\documentclass[pra,twocolumn,showpacs,floatfix,letterpaper,superscriptaddress,longbibliography]{revtex4-2}

\usepackage[utf8]{inputenc}
\usepackage[T1]{fontenc}

\usepackage{amsmath}
\usepackage{amsfonts}
\usepackage{amssymb}
\usepackage{newtxtext,newtxmath}
\usepackage{physics}    

\usepackage{ragged2e}
\usepackage{subcaption}
\usepackage{graphicx}
\usepackage{epstopdf}
\usepackage[percent]{overpic}
\usepackage{float}

\usepackage{xcolor}
\usepackage{comment}

\usepackage[colorlinks=true,citecolor=blue,linkcolor=blue,urlcolor=blue]{hyperref}

\begin{document}	
	
	\title{Shot-based variational simulation of the toric code phase transition: \\
		Energy, global entanglement, and noise resilience}
	\author{Hayede Zarei}
	\email{s.hayede.zarei@gmail.com}
	\affiliation{Physics Department, College of Sciences, Shiraz University, Shiraz 71454, Iran}
	\author{Mohammad Hossein Zarei}
	\email{mzarei92@shirazu.ac.ir}
	\affiliation{Physics Department, College of Sciences, Shiraz University, Shiraz 71454, Iran}
	
	\date{\today}
	
	\begin{abstract}
		We investigate the capability of variational quantum circuits in detecting a topological phase transition within the toric code model subjected to a uniform magnetic field. The detection is carried out via shot-based computations of both the ground state energy and the global entanglement. We begin by employing well-established renormalization techniques to construct a quantum circuit that accurately represents the pure toric code under periodic boundary condition. This circuit is subsequently parameterized to enable a shot-based variational simulation of the ground state in the presence of the magnetic field. Our results demonstrate that even for the minimal lattice sizes and a finite number of measurement shots, the ground state energy exhibits a distinct singularity at the transition point which can be reliably approximated through finite-size scaling analysis. Furthermore, we introduce a secondary variational circuit designed to compute the global entanglement of the generated ground state. We observe that both global entanglement and its conditional counterpart display the expected qualitative behavior at the transition point, consistent with the topological nature of the phase transition. To evaluate the practical viability of our variational approach on real quantum hardware, we systematically examine the detrimental effects of gate noise on the analysis. We find that the topological phase exhibits a notable robustness against noise, manifesting as a linear decrease in the transition point with increasing noise strength.     
	\end{abstract}
	
	\maketitle

	\section{Introduction}


Identifying quantum phases of matter is a central problem in condensed matter physics~\cite{sachdev2023quantum}. In particular, quantum phases with topological order are of great interest because of their non-local nature, which cannot be characterized by any local order parameter~\cite{wen1990topological,Wen1,Wen2,levin2006detecting,wen1990topological,kitaev2003}. Since accurately computing the properties of correlated quantum systems is computationally demanding for classical methods~\cite{troyer2005computational}, simulating such systems such as topological phases on real quantum computers has attracted much attention~\cite{park2016simulation,fossfeig2023adaptive,xu2024digital,semeghini2021probing,barreiro2011open}. Moreover, recent advances in the development of quantum processors across various platforms, including superconducting qubits~\cite{arute2019quantum} and Rydberg atom arrays~\cite{ebadi2021quantum}, have made it possible to study and simulate such phases. These systems are now capable of hosting hundreds of physical qubits, and their scale is expected to increase significantly in the near future~\cite{ebadi2021quantum,browaeys2020manybody}.

Among different approaches for simulating topological quantum systems on quantum computers~\cite{altman2021quantum,shi2023quantum}, variational quantum algorithms---most notably the variational quantum eigensolver (VQE)---play an important role~\cite{peruzzo2014variational,mcclean2016theory,zhang2022digital,zapletal2024error}. They provide an efficient framework for approximating the ground state of local Hamiltonians on quantum processors. In this approach, the system's wave function is prepared as a parameterized quantum state (ansatz), and the optimal approximation to the ground state is obtained by minimizing the expectation value of the Hamiltonian. Given the feasibility of implementing such algorithms on near-term quantum computers, an important part of studies on quantum simulators has been focused on variational simulation of quantum systems~\cite{kandala2017hardware,nam2020ground,cerezo2021variational,yuan2019theory,lyu2023symmetry,lyu2020accelerated}. In particular, since topological quantum states have a complex structure of quantum correlations including long-range entanglement~\cite{chen2010local}, finding a suitable variational quantum circuit is a challenging task~\cite{ciaramelletti2025detecting,sim2019expressibility,tang2021qubit,du2020expressive,mcclean2018barren}. In particular, it is important to ensure that a parameterized variational circuit captures the long-range entanglement for some values of variational parameters in the sense that one is able to identify a topologically ordered phase in a Hamiltonian and to probe its properties. 

One of the most well-known topological systems is toric code which plays a paradigmatic role for studying topological order~\cite{kitaev2003}. It is important both from a fundamental perspective---due to features such as ground-state topological degeneracy, anyonic excitations, and a finite energy gap~\cite{exp,homeier2021z}---and from a practical perspective because of its role in quantum error correction~\cite{dennis2002}. A detailed study of this model therefore provides key insights into the physics of topological phases~\cite{Levin,bravyi2010topological,araujoderesende2020}. In particular, toric code provides a natural platform for applying VQE and probing physical properties associated with topological phases~\cite{iqbal2024topological,aktar2026quantum}. In this regard, quantum circuits for preparing the ground state of the toric code model have been proposed ~\cite{chen2024quantumcircuits,liu2022methods,fowler2012surface,piroli2021quantum} and, experimentally implemented in recent years~\cite{satzinger2021realizing}.

Studying on the toric code model is specifically important for understanding nature of topological phase transitions where toric code Hamiltonian is perturbed by some local perturbations~\cite{trebst2007breakdown,dusuel2011robustness}. One of the best-known models is the toric code in the presence of a magnetic field where the topological phase of toric code is robust against string tension and the system undergoes a transition to a polarized phase at a critical magnetic field \cite{trebst2007breakdown}. This phase transition shows signatures in different quantities such as topological entanglement entropy~\cite{hamma2005topological}, ground state fidelity~\cite{zanardi2006ground} and string order parameter~\cite{zarei2019ising}. Regarding these important features, toric code in presence of a magnetic field is a key candidate for simulation of topological phase transitions on a quantum computer.  In particular, recently a parameterized quantum circuit has been proposed for state-vector-based simulation of this model~\cite{sun2023parametrized}.

On the other hand, in order to mimic a real quantum computer, one should use a shot-based simulator where the state vector is not available and properties of the ground state should be characterized by analyzing outcome of a finite number of measurement shots~\cite{huang2020predicting,peruzzo2014variational,mcclean2016theory}. In particular, finite sampling can introduce fluctuations comparable to, or even larger than, the physical variations of interest, obscure the optimizer's actual direction of improvement, and affect the location or sharpness of features associated with the phase transition~\cite{bravyi2021mitigating,gard2020efficient,li2024ensemble}. Therefore, it is an important task to  obtain stable results with a finite shot budget~\cite{sweke2020stochastic}. This demonstrates that the proposed circuit not only possesses adequate representational capability in the ideal limit, but is also efficient in terms of measurement overhead, optimization stability, and the statistical extraction of physical information. 

In this regard, we perform a shot-based simulation of the toric code phase transition using the QASM Simulator implemented in Qiskit Aer~\cite{abraham2019qiskit,cerezo2021variational}. We show that by considering the smallest possible sizes of the lattice and in spite of a finite number of shots, one is able to capture thermodynamic limit of the problem by using finite-size scaling analysis. Since considering periodic boundary condition for toric code plays an important role in analyzing thermodynamic limit, we first develop a step-by-step method for designing a quantum circuit which generates toric code state on a torus with various system sizes. Then we extend the above circuit to a parameterized version, as proposed in Ref.~\cite{sun2023parametrized}, which serves as a variational quantum circuit for finding ground state of toric code in magnetic field. We consider a finite number of measurement shots to compute the ground-state energy for different lattice sizes, and consider signatures of the quantum phase transition by analyzing the derivatives of the energy and using the finite-size scaling method to estimate the transition point. This analysis enables comparison of the obtained results with the known behavior of the toric code model and thus provides a benchmark for assessing the performance of the proposed framework.

Moreover, we notice that regarding rich structure of entanglement in topological phases, it is also important to characterize topological phase transitions by studying different measures of entanglement~\cite{hamma2008entanglement,hamma2005ground}. Although topological entanglement entropy is the most well-known quantity~\cite{kitaev2006topological}, it is expected that measures of multi-partite entanglement can show signatures of a phase transition~\cite{de2006multipartite,de2006global,montakhab2010multipartite}. In particular, global entanglement and conditional global entanglement have been recently considered for the toric code in presence of a nonlinear perturbation where these quantities show meaningful signature of topological phase transition~\cite{samimi2022global,samimi2023conditional}. Since outcome of our variational circuit approximates the ground state of the toric code in a magnetic field, we are also able to compute similar quantities for our model. To this end, we introduce a second variational circuit at the end of the first circuit to find global entanglement and conditional global entanglement and, accordingly, we plot these quantities as a function of the magnetic field. We show that despite the small size of the lattice and finite number of measurement shots, our results are in very good agreement with the results derived in~\cite{samimi2022global} where conditional global entanglement shows a peak at the transition point.

Finally, we notice that, in a real quantum computer, the effect of noise cannot be ignored~\cite{barron2021preserving,preskill2018quantum} and therefore, it is also important to incorporate a realistic noise model in our simulation. Specifically, we apply one- and two-qubit depolarizing noise channels after the corresponding quantum gates in the variational circuit, and also account for readout errors at the measurement stage~\cite{wallman2016noise,bravyi2021mitigating}. We then reconsider our analysis of ground-state energy to study the effect of noise in the characterization of the transition point. We show that, for small values of noise probability, the transition point shifts linearly with increasing noise. This linear reduction implies that topological phase exhibits robustness against noise and therefore, topological phase transitions are good candidates for simulation on noisy quantum computers.

The remainder of this paper is organized as follows. Sec. \ref{sec2} presents a brief review of the toric code model. Sec. \ref{sec3} introduces the structure of the quantum circuit used to prepare the ground state of this model. Sec. \ref{sec4} investigates the behavior of the system in the presence of a magnetic field; in this section, we introduce the framework of the variational quantum eigensolver (VQE) and present the method for calculating the ground-state energy and global entanglement. Sec. \ref{sec5} studies the effect of realistic noise on the circuit performance and the resulting measurement outcomes. Finally, Sec. \ref{sec6} provides a summary and concluding remarks.

	\section{Toric Code Model}\label{sec2}
	
	The toric code is a paradigmatic model of topological order and a cornerstone of fault-tolerant quantum computation. It is defined on a square lattice with periodic boundary conditions, corresponding to a lattice embedded on the surface of a torus, where qubits reside on the edges of the lattice. The dynamics of the system are governed by a Hamiltonian composed of mutually commuting local stabilizers:
	\begin{equation}
	\mathcal{H}_{\mathrm{TC}} = -\sum_s A_s - \sum_p B_p .
	\end{equation}
	
\begin{figure}[h!]
	\centering
    \includegraphics[width=8.7cm,height=5.2cm,angle=0]{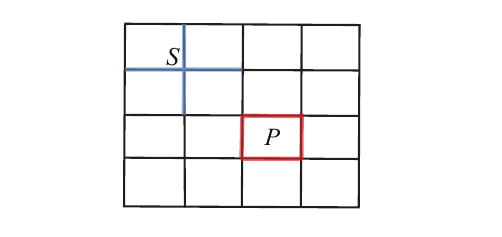}
	\caption{Square-lattice geometry of the toric code, showing a vertex ($S$) and a plaquette ($P$).}
	\label{fig:toric-code}
	\end{figure}

		Here $A_s$ and $B_p$ denote the vertex and plaquette operators, respectively, defined as
	\begin{equation}
		A_s = \prod_{i \in s} Z_i , \qquad
		B_p = \prod_{j \in p} X_j ,
	\end{equation}
	where $X$ and $Z$ are Pauli operators, $i \in s$ refers to qubits incident to a vertex labeled by $s$ and $j \in p$ refers to qubits around a plaquette labeled by $p$, see Fig.~\ref{fig:toric-code}.
	
\begin{figure}[h]
		\centering
		\includegraphics[width=8.7cm,height=4.5cm,angle=0]{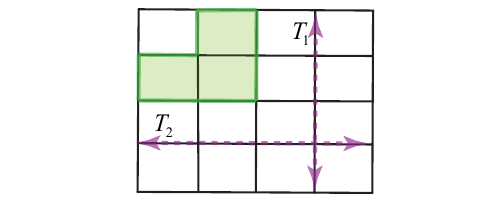}
		\caption{Loop representation in the toric code, showing a closed plaquette loop and the noncontractible loops associated with $T_1$ and $T_2$.}	
		\label{fig:toric-loop}
	\end{figure}

	Since all vertex and plaquette stabilizers commute with one another, the model is exactly solvable and is placed in the class of frustration-free systems. The ground state $|\Psi_{00}\rangle$ is the simultaneous $+1$ eigenstate of all stabilizers. It can be written as an equal superposition of all closed-loop configurations, obtained by applying the plaquette projectors to the fully polarized reference state:
	\begin{equation}
		|\Psi_{00}\rangle =
		\prod_{p=1}^{N_p -1}
		\frac{1}{\sqrt{2}}\left(I_p + B_p\right)
		|00\ldots0\rangle ,
	\end{equation}
		where $I_p$ refers to an Identity operator corresponding to plaquette $p$ and $N_p$ is the number of plaquettes of the lattice. Due to periodic boundary conditions, it follows that $\prod_{p=1}^{N_p} B_p =1$ and therefore, in the above equation  $\prod_{p=1}^{N_p -1}$ refers to product of all $N_p -1$ independent plaquette operators. On the other hand, as shown in Fig.~\ref{fig:toric-loop}, a product of $B_p$ operators can be represented by a loop in the lattice. In this regard, the ground state of the toric code can be interpreted as a uniform superposition of closed-loop configurations and therefore, it is called a loop-gas model.
	
	Since the lattice is defined on a torus, the ground state exhibits a fourfold degeneracy that originates from the global topology of the manifold. Transitions between these ground states cannot be generated by any local operator. Instead, they require nonlocal string operators that follow noncontractible loops around the torus. As shown in Fig.~\ref{fig:toric-loop}, nonlocal operators $T_1$ and $T_2$ are defined by the products of $X$ operators over qubits residing on two noncontractible loops around the torus, respectively. In this regard, other ground states of the model are constructed in the following form:
	\begin{equation}
|\Psi_{ij}\rangle =	T_1^i T_2^j	|\Psi_{00}\rangle ,
	\end{equation}
	where $i$ and $j$ take binary values. An important point is that as the lattice size $L \times L$ increases, the length of these nonlocal strings grows. Accordingly, it is shown that the logical error rate becomes exponentially suppressed with increasing system size~\cite{dennis2002}. This property underlies the use of the toric code as a topologically protected quantum memory. In particular, notice that topological degeneracy is also robust against local perturbations added to the Hamiltonian provided the system remains in the topological phase for small perturbation strengths and undergoes a topological phase transition at a critical value of the perturbation power~\cite{bravyi2010topological,kitaev2003}. 
	
    Regarding the above important physics behind topological order in the toric code, simulation of toric code state on a quantum computer is very important for studying properties of topological quantum phases. Such simulations make it possible to consider which order parameter or quantum information quantity such as multipartite entanglement can probe the nature of a topological phase. In this regard, introducing a suitable quantum circuit which generates a toric code state is a key step. In the following section, we introduce a scalable circuit construction strategy that enables the preparation of the toric code ground state.
	
	\section{Quantum Circuit for pure Toric code }\label{sec3}
Various studies have focused on designing quantum circuits for toric code state ~\cite{fowler2012surface,piroli2021quantum}. In particular, the toric code model with periodic boundary conditions is more challenging as it requires either nonlocal connections or more complex circuit constructions to emulate periodic boundary conditions. Recently, Chen and co-workers~\cite{chen2024quantumcircuits} proposed a method for constructing a quantum circuit for the $L \times L$ toric code model on a torus, in which the ground state is prepared using a circuit of depth $2L + 2$ through the sequential application of Hadamard and CNOT gates. However, as also noted in that work, when the lattice is placed on more complex or asymmetric geometries, determining the appropriate ordering of the operators becomes a subtle issue, and consequently extending this method to lattices larger than $3 \times 3$ or to more general geometries is challenging. For this reason, a second approach, referred to as the Gluing method, has also been proposed. This method, which combines Clifford gates with measurements of the Pauli $X$ operator, yields circuits with a smaller depth than the first approach. Nevertheless, since repeated measurements entail a significant execution cost on current quantum processors, such an approach is not particularly favorable for experimental implementation. 

In this section, we start from toric code on a $2\times 2$ lattice on a torus and consider quantum circuit proposed in~\cite{chen2024quantumcircuits} which is based on applying a sequence of Hadamard and CNOT gates. Then, we introduce a step-by-step strategy to extend the above circuit to arbitrary $L\times L'$ lattice with periodic boundary condition. Our strategy is based on the concept of lattice expansion, which originates from triangulation transformations of lattice~\cite{dennis2002}. Vidal~\cite{vidal2008entanglement} employed this framework in the context of renormalization to reduce the effective system size. Here, we adopt a different perspective and reinterpret it as a constructive algorithm for state synthesis on quantum hardware. To this end, we notice that stabilizer formalism provides a convenient framework for tracking operator action in surface-code systems~\cite{dennis2002}. Within this formalism, elementary moves are defined as local transformations that modify the lattice and its associated qubits while preserving the topological structure of the code. In general, these moves involve the addition or removal of vertices and plaquettes together with their corresponding qubits. In particular, we consider two simple steps as follows. 
	
\begin{figure}[t]
	\centering
	\includegraphics[width=1\linewidth]{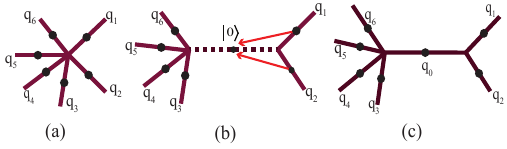}
	\caption{Schematic illustration of vertex addition in the toric code. (a) Initial vertex with six edges. (b) Integration of the ancillary qubit $q_0$ via CNOT gates from $q_1$ and $q_2$ to $q_0$ to form the new vertex stabilizer. (c) The resulting expanded lattice structure with the new vertex.}
	\label{fig:vertex_addition}
\end{figure}

	\textbf{Vertex inserting:} For example, consider toric code on a specific lattice with a vertex connected to six edges. The corresponding stabilizer operator is $Z_1 Z_2 Z_3 Z_4 Z_5 Z_6$. To add a vertex to the lattice, an ancillary qubit $q_0$, initialized in the state $|0\rangle$ is introduced into the system as shown in Fig.~\ref{fig:vertex_addition}. Then a sequence of CNOT gates is applied from the neighboring qubits $q_1$ and $q_2$ with this new qubit as the target. In this representation, the tail and head of each arrow denote the control and target qubits, respectively. It is known that when a CNOT gate is applied, with the first qubit as the control, to an operator of the form $I \otimes Z$, it is transformed into $Z \otimes Z$.
	 Accordingly, by applying CNOT gates between qubits $q_1$ ($q_2$) and $q_0$ as shown in Fig.~\ref{fig:vertex_addition}b, initial stabilizer of $Z_0$ is transformed to a new vertex stabilizer of the form $Z_0 Z_1 Z_2$. Moreover, the product of this operator with the original stabilizer $Z_1 Z_2 Z_3 Z_4 Z_5 Z_6$ yields another stabilizer operator $Z_0 Z_3 Z_4 Z_5 Z_6$, which is consistent with the stabilizer structure of toric code on the expanded lattice Fig.~\ref{fig:vertex_addition}c.

\begin{figure}[t]
	\centering
	\includegraphics[width=1
	\linewidth]{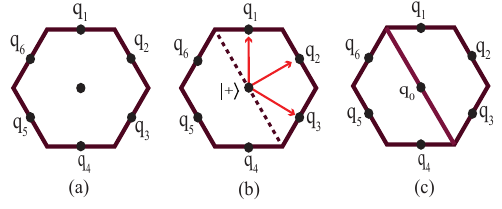}
	\caption{Schematic illustration of edge addition in the toric code. (a) A hexagonal lattice to which an additional edge is to be added. (b) Unidirectional CNOT gates from the ancillary qubit in the $|+\rangle$ state to all neighboring qubits. (c) The resulting lattice after the edge addition.}
	\label{fig:plaquette_addition}
\end{figure}

		\textbf{Face adding:} To add a plaquette to the initial lattice, we apply a symmetric operation that is dual to the vertex-addition procedure. For example, consider toric code on a lattice with a hexagonal plaquette. Then, an ancillary qubit initialized in the state $|0\rangle$ is introduced and then is transformed to the state $|+\rangle$ by applying a Hadamard gate. However, unlike the vertex case, the CNOT gates are applied from the new qubit, as the control qubit, to its neighboring qubits $q_1$, $q_2$ and $q_3$ within the plaquette.  It is known that under the action of a CNOT gate, with the first qubit as the control qubit, on an operator in the form of $X\otimes 1$, it is transformed into $X\otimes X$. Accordingly, by applying CNOT gates between qubits $q_0$ and $q_1$ ($q_2$, $q_3$) as shown in Fig.~\ref{fig:plaquette_addition}b, the initial stabilizer $X_0$ is transformed into a new plaquette stabilizer of the form $X_0 X_1 X_2 X_3$. Moreover, product of this operator with the original stabilizer yields another stabilizer operator $X_0 X_4 X_5 X_6$ which corresponds to another new plaquette (Fig.~\ref{fig:plaquette_addition}c).

\begin{figure*}[t]
	\centering
	\includegraphics[width=\textwidth]{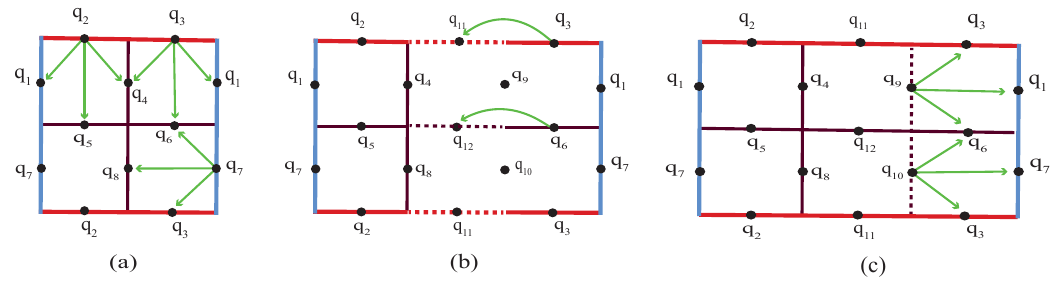}
	\caption{Extension of the Toric Code lattice from $2 \times 2$ to $2 \times 3$. 
		(a) Initial $2 \times 2$ lattice with periodic boundary conditions, where CNOT gates are applied to generate the initial plaquette stabilizers. 
		(b) Vertex insertion: introduction of ancillary qubits $q_{11}$ and $q_{12}$ (initialized in $|0\rangle$) and their entanglement via CNOT gates to expand the lattice horizontally. 
		(c) Plaquette addition: introduction of qubits $q_9$ and $q_{10}$ (initialized in $|+\rangle$) and application of CNOT gates to complete the $2 \times 3$ lattice structure.}
	\label{fig:lattice_extension}
\end{figure*}

By the above simple deformation, we are able to construct the toric code on a square lattice with arbitrary size. To this end, we start with the smallest nontrivial instance of the toric code, namely a $2 \times 2$ lattice on a torus with periodic boundary conditions. As shown in Fig.~\ref{fig:lattice_extension}a, we initialize all qubits in the state $|0\rangle^{\otimes n}$. Notice that in the figure, the edges on the top and bottom sides of the lattice, as well as those on the right and left sides, are identified according to the periodic boundary conditions. Hadamard gates are applied to a selected subset of edges, specifically $q_2$, $q_3$, and $q_7$ in Fig.~\ref{fig:lattice_extension}a, in order to provide the initial stabilizers $X_2$, $X_3$, and $X_7$. Subsequently, CNOT gates are applied according to the directed pattern shown in Fig.~\ref{fig:lattice_extension}a between qubits $q_2$, $q_3$, and $q_7$ and their corresponding neighbors. In this regard, the initial stabilizers are transformed into plaquette operators of the toric code state.

Extending the $2 \times 2$ lattice to an arbitrary lattice size, is sufficient to enlarge the toric code using the vertex-insertion and face-adding transformations. As shown in Fig.~\ref{fig:lattice_extension}b, two new qubits, $q_{11}$ and $q_{12}$, initialized in the state $|0\rangle$, are inserted on the edges of the initial lattice and entangled with the existing network through the corresponding CNOT operations. As a result, a new lattice with additional vertices is obtained, where the new vertices are inserted between $q_3$ ($q_6$) and $q_{11}$ ($q_{12}$). In the second step, the qubits $q_{10}$ and $q_9$, prepared in the state $|+\rangle$, are added to the initial faces of the lattice, as shown in Fig.~\ref{fig:lattice_extension}c. Then, by applying the appropriate CNOT operations, the additional plaquettes are generated, yielding the toric code on a $2 \times 3$ lattice. By repeating the same procedure, one can add further cells along both lattice directions and thus enlarge the lattice to the arbitrary size. 

In this way, one obtains a lattice-based representation of the pattern of Hadamard and CNOT gates required to construct a toric code state on a square lattice with arbitrary size. This pattern can be translated straightforwardly into circuit language for implementation on a quantum computer. In Fig.~\ref{circuit_0}, we show the corresponding circuit for the toric code on a $2 \times 2$ lattice. In the next section, we use this method to construct quantum circuit corresponding to $3\times 3$ and $4\times 4$ lattices for varitional simulation of the toric code phase transition.  

	\section{Toric Code in the Presence of a Magnetic Field: A variational approach}\label{sec4}
	Quantum circuit for generating a toric code state is a base for studying non-solvable versions of toric code model by variational quantum algorithms. An important example is toric code in the presence of a uniform magnetic field applied along the $z$-direction,
	\begin{equation}
			\mathcal{H} = - \sum_s A_s -  \sum_p B_p - h \sum_i Z_i .
	\end{equation}
		Here the last term refers to the magnetic field of strength $h$.
		
		When $h=0$ reduces to the pure toric code Hamiltonian and the ground state exhibits a non-trivial topological order. However, when the magnetic field increases, the third term overcomes and the ground state is a fully polarized state $|00...0\rangle$ which is a topologically trivial phase. In this regard, it is clear that the above Hamiltonian shows a topological phase transition which occurs at a critical field whose value is known to be $h_c = 0.328$~\cite{trebst2007breakdown}. This model has been extensively studied in the literature as a toy model of topological phase transition. Consequently, quantum simulation of such a model on a quantum computer is an important task for studying capability of quantum computers in characterization of topological quantum order. 

	\subsection{Simulation using a Variational Quantum Eigensolver (VQE)}
	To simulate the ground state of the toric code in the presence of a magnetic field, we employ the Variational Quantum Eigensolver (VQE) algorithm. For this purpose, we use the idea presented in \cite{sun2023parametrized} to construct a parameterized quantum circuit capable of capturing the dominant entanglement structure of the ground state across different magnetic field regimes. We recall that the ground state of the pure toric code is an equal-weight superposition of closed-loop configurations generated by products of the star ($A_s$) and plaquette ($B_p$) operators. However, because the magnetic field term does not commute with the plaquette operators, the true ground state becomes a weighted superposition of loop states. 
	
	Physically, the magnetic field modifies the relative probability amplitudes of different loop configurations. Consequently, a weighted loop-gas state serves as a highly suitable variational ansatz for capturing the ground-state physics of the perturbed toric code. A simple approach to constructing such a variational state is to modify the pure toric code preparation circuit by replacing the Hadamard gates (H) with single-qubit rotation operators around $y$ axis $U(\theta) = R_y(\theta)$, defined as:
	\begin{equation}
		U(\theta) = 
		\begin{pmatrix}
			\cos(\theta/2) & \sin(\theta/2) \\
			-\sin(\theta/2) & \cos(\theta/2)
		\end{pmatrix}.
	\end{equation}

These rotations act on the reference state as $U(\theta)|0\rangle = \cos(\theta/2)|0\rangle + \sin(\theta/2)|1\rangle$, which produces a weighted superposition of the computational basis states. The parameter $\theta$ directly controls the relative amplitudes of the $|0\rangle$ and $|1\rangle$ components, generating a continuously tunable family of real-valued states. This construction connects smoothly to the zero-field limit where $\theta = \pi/2$ and $U(\theta)$ reduces to the standard Hadamard transformation, thereby recovering the exact circuit for the pure toric code. On the other hand, for $\theta=0$, $U(\theta)$ is an Identity operator and therefore the outcome of the circuit would be a polarized state $|000...0\rangle$ corresponding to infinite-field limit of the Hamiltonian.

The replacement $H \rightarrow U(\theta)$ can thus be viewed as a minimal parametric extension of the circuit preparing the ground state. Deviations of $\theta$ from $\pi/2$ adjust the relative weights of loop configurations in response to the magnetic field. Consequently, the resulting circuit provides a simple variational ansatz for studying the toric code in a magnetic field.

We now employ the VQE algorithm to determine the ground state. For the parameterized state $|\psi(\theta)\rangle$ generated by the quantum circuit, the expectation value of the Hamiltonian is evaluated as:
\begin{equation}
	E(\theta) = \langle\psi(\theta)| \mathcal{H} |\psi(\theta)\rangle.
\end{equation}

Since the Hamiltonian is constructed from Pauli operators, the expectation value above can be estimated efficiently by measuring the corresponding Pauli terms. The optimal parameter $\theta^*$ is obtained by minimizing the energy with respect to $\theta$, thereby yielding the best variational approximation to the ground state within the proposed ansatz. Note that all energy evaluations were performed using the QASM simulator, where the expectation values of the Hamiltonian terms were estimated by shot-based sampling of the circuit outputs. As a result, unlike exact statevector simulation, the data exhibit statistical fluctuations due to the finite number of measurement shots.

\begin{figure}[h]
	\centering
	\includegraphics[width=8.7cm,height=5.2cm,angle=0]{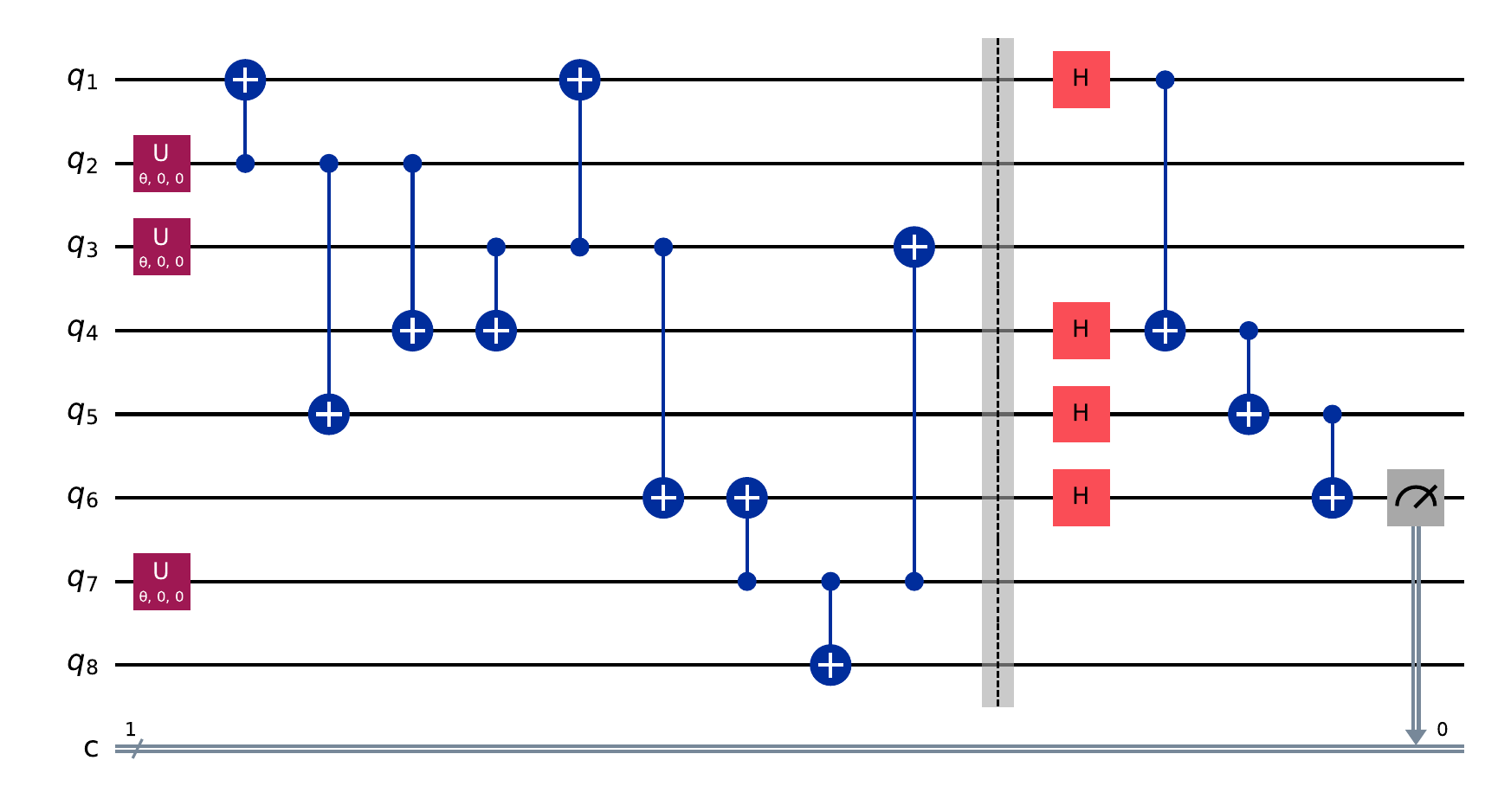}
	\caption{Quantum circuit for generating the variational ground state corresponding to the toric code on $2\times 2$ lattice in Fig.\ref{fig:lattice_extension}-a. By replacing rotation operators $U$ with Hadamard gates, the circuit generates the pure toric code. The second part of the circuit represents measurement circuit corresponding to a plaquette operator $B_p = X_1X_4X_5X_6$.}
	\label{circuit_0}
\end{figure}

As an example, in Fig.~\ref{circuit_0} we show the quantum circuit used to measure the plaquette operator $B_p$ in the toric code on a $2 \times 2$ lattice. To this end, the first part of the circuit implements a variational ansatz to prepare a state that approximates the ground state of the toric code in the presence of a magnetic field. The second part of the circuit is then used to evaluate the plaquette term. Specifically, Hadamard gates are applied to rotate the measurement basis of the relevant qubits into the $X$ basis, and a sequence of CNOT gates is used to entangle the four physical qubits such that the parity information of the $B_p$ operator is mapped onto one of them. Consequently, the expectation value of the plaquette operator can be extracted from a single-qubit measurement. This procedure is repeated to evaluate the expectation values of the different Hamiltonian terms, from which the energy $E(\theta)$ is obtained. Finally, by comparing the energies computed for different values of $\theta$, the parameter corresponding to the minimum energy is selected as the optimal result.

\begin{figure*}[t]
	\centering
	\begin{subfigure}[b]{0.32\textwidth}
		\centering
		\includegraphics[width=\linewidth]{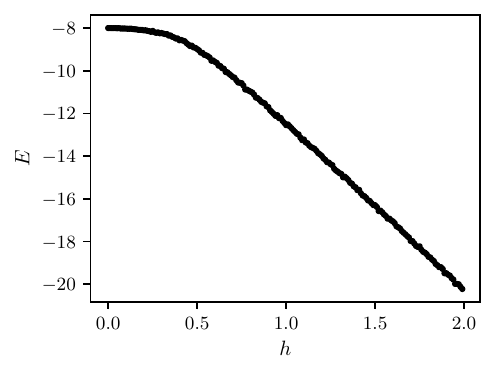}
		\caption{}
		\label{fig:8q_energy}
	\end{subfigure}
	\hfill
	\begin{subfigure}[b]{0.32\textwidth}
		\centering
		\includegraphics[width=\linewidth]{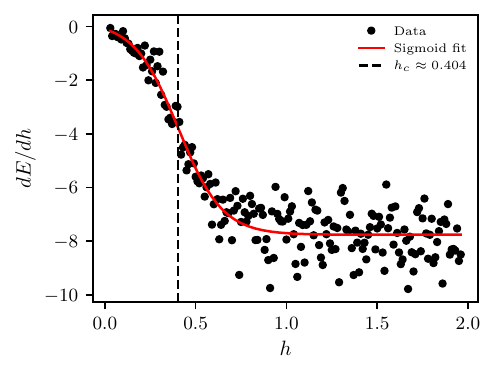}
		\caption{}
		\label{fig:8q_energy_derivative}
	\end{subfigure}
	\hfill
	\begin{subfigure}[b]{0.32\textwidth}
		\centering
		\includegraphics[width=\linewidth]{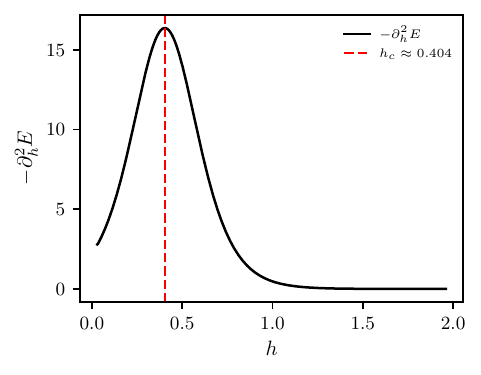}
		\caption{}
		\label{fig:8q_energy_second_derivative}
	\end{subfigure}
	
	\caption{Ground-state energy and its derivatives for the toric code on a $2 \times 2$ lattice, calculated using 1024 measurement shots. (a) Energy as a function of the magnetic-field strength $h$. (b) Numerical derivative $dE/dh$ fitted with a sigmoid function (red line) to estimate the transition point $h_c \approx 0.404$. (c) Second derivative $-\partial_h^2 E$ derived from the fit, highlighting the critical region.}
	\label{fig:combined_8q_plots}
\end{figure*}

The ground-state energy is computed for different values of the magnetic-field strength $h$. In Fig.~\ref{fig:8q_energy}, we show the ground-state energy as a function of $h$ in the interval $0 \leq h \leq 2$, sampled at 200 points.

To identify signatures of a phase transition, the numerical derivative $dE/dh$ was computed using a central-difference scheme with a suitably chosen step size to balance numerical precision against sampling fluctuations (Fig.~\ref{fig:8q_energy_derivative}). While these fluctuations are small in the energy itself, they become more pronounced when numerical derivatives are evaluated. Despite finite-size effects and sampling noise, the derivative exhibits a clear transition-like behavior where it changes from values close to zero in the topological phase to large negative values in the polarized phase. A smooth sigmoid function fitted to the data yields an estimated transition point $h_c \approx 0.404$, slightly larger than the thermodynamic-limit value $h_c \approx 0.33$--$0.34$, as expected for a system of finite size.

	For a clearer visualization of the transition region, we consider the quantity $-\partial_h^2 E$, obtained from the analytical derivative of the fitted sigmoid function (Fig.~\ref{fig:8q_energy_second_derivative}). This quantity exhibits a pronounced peak near the transition. In many-body systems, such a peak reflects the maximal rate of change in ground-state correlations and can be interpreted as a susceptibility-like response near the critical point~\cite{SachdevQuantumPhaseTransitions}. As the system size increases, this feature is expected to sharpen, eventually diverging as a true singularity in the thermodynamic limit.

\begin{figure*}[t]
	\centering
	\begin{subfigure}[t]{0.32\textwidth}
		\centering
		\includegraphics[width=\linewidth]{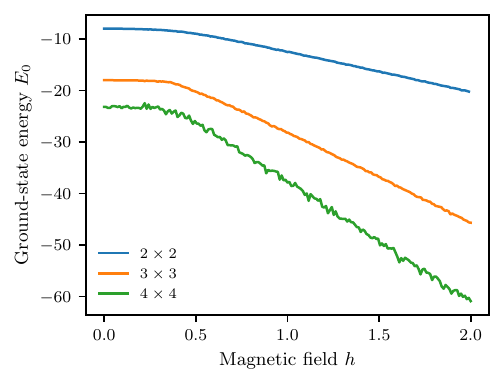}
		\caption{}
		\label{fig:comparison_energy}
	\end{subfigure}
	\hfill
	\begin{subfigure}[t]{0.32\textwidth}
		\centering
		\includegraphics[width=\linewidth]{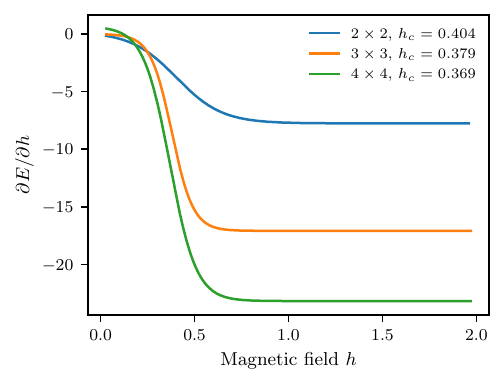}
		\caption{}
		\label{fig:comparison_derivative}
	\end{subfigure}
	\hfill
	\begin{subfigure}[t]{0.32\textwidth}
		\centering
		\includegraphics[width=\linewidth]{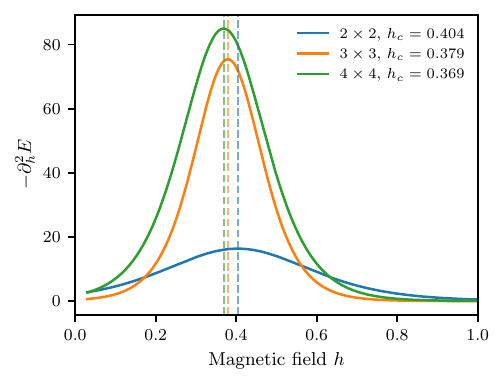}
		\caption{}
		\label{fig:comparison_second_derivative}
	\end{subfigure}
	
	\caption{Comparison of the ground-state energy and its derivatives for different lattice sizes ($2 \times 2$, $3 \times 3$, and $4 \times 4$). (a) Ground-state energy as a function of the magnetic field $h$. The increased fluctuations in the larger systems arise from the reduced measurement budget where 1024 shots were used for the $2 \times 2$ lattice, 512 shots for the $3 \times 3$ lattice, and 32 shots for the $4 \times 4$ lattice. (b) First derivative of the energy, $dE/dh$, illustrating the sharpening of the transition region with increasing system size. (c) Analytical second derivative, $-\partial^2 E / \partial h^2$, obtained from the sigmoid fits, where the sharpening of the peak reflects the developing singularity near the critical point.}
	\label{fig:comparison_3_lattices}
\end{figure*}

	To estimate transition point in thermodynamic limits we should use a finite-size scaling approach. To this end, we need to apply the same VQE protocol for other system sizes. Here we consider $3\times3$ and $4\times4$ lattices and then compare all system sizes within a unified set of plots. For the larger lattices, the number of measurement shots was reduced due to computational constraints. As a result, the data for the $3\times3$ system exhibit noticeably stronger statistical fluctuations, which are already visible in the corresponding energy curve, Fig.~\ref{fig:comparison_3_lattices}a.
	
	Despite these fluctuations, the comparison across system sizes reveals a consistent trend where the transition region in the energy becomes sharper as the lattice grows. This behavior is echoed in the first derivative $dE/dh$, where the variation near the transition becomes increasingly pronounced for larger systems. The marked points on the curves indicate the extracted critical fields for each lattice size $L$ and show that $h_c(L)$ moves steadily toward the expected thermodynamic limit value as $L$ increases, Fig.~\ref{fig:comparison_3_lattices}b.
	
	A similar scaling behavior is observed in Fig.~\ref{fig:8q_energy_second_derivative} for the quantity $-\partial_h^2 E$, obtained from the analytical derivative of the sigmoid fit to $dE/dh$. The peak height increases with system size, reflecting a stronger response of the system in the transition region and a gradual suppression of finite-size effects(Fig.~\ref{fig:comparison_3_lattices}c). This sharpening of the peak is consistent with the approach to the thermodynamic limit, where the susceptibility develops an actual singularity at the critical point.
	
\begin{figure}[h]
		\centering
		\includegraphics[width=\linewidth]{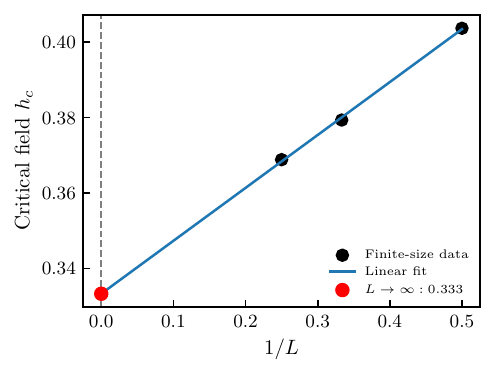}
	\caption{Finite-size scaling of the critical field $h_c$ versus $1/L$. The linear fit extrapolates the VQE results (blue circles) to the thermodynamic limit at $1/L=0$ (red circle).}
	
		\label{fig:8q_finite_size_scaling}
	\end{figure}

	To estimate the critical point in the thermodynamic limit, the extracted values $h_c(L)$ for different lattice sizes were plotted as a function of $1/L$, where $L$ denotes the linear system size ($L=2, 3, 4$ for the $2\times2$, $3\times3$, and $4\times4$ lattices, respectively), as shown in Fig.~\ref{fig:comparison_3_lattices}. Following the standard finite-size scaling ansatz, a linear fit in $1/L$ was performed on these data. Despite the limited number of system sizes accessible in our simulations, the data exhibit a clear linear trend within numerical uncertainty (see Fig.~\ref{fig:8q_finite_size_scaling}). Extrapolating the fit to the limit $1/L \rightarrow 0$ yields $h_c^{(\infty)} \approx 0.333$. This value is consistent with the expected theoretical result, indicating that the deviations of $h_c(L)$ observed for small lattices arise predominantly from finite-size effects.
	
	These results indicate that even the smallest non-trivial instances of the model already capture qualitative signatures of the transition from the topological phase to the polarized phase. The VQE approach thus provides a practical and scalable framework for exploring the breakdown of topological order in perturbed stabilizer Hamiltonians. 

	\subsection{Characterizing topological phase transition by global entanglement}
	Quantum phase transitions are generally associated with changes in multipartite entanglement. In topological phase transitions, a particularly relevant quantity is the topological entanglement entropy \cite{kitaev2006topological,levin2006detecting}, which exhibits a discontinuous jump at the transition point in the toric code in the presence of a magnetic field \cite{hamma2005topological}. However, other entanglement measures also provide valuable information about the nature of topological phase transitions. For the toric code in a magnetic field, several such measures have been studied, whereas global entanglement (GE) has received comparatively little attention. In particular, the response of the toric code to a non-linear perturbation has been investigated in related settings~\cite{samimi2022global}. In this work, we examine GE for the toric code in a magnetic field and compare our results with those of previous studies.
	
GE is defined as the mean linear entropy of the single-qubit reduced density matrices of an $N$-qubit many-body state,
\begin{equation}
	GE = \frac{1}{N} \sum_i \left( 1 - \operatorname{tr}(\rho_i^2) \right),
\end{equation}
where $\rho_i$ denotes the reduced density matrix of the $i$th qubit~\cite{MeyerWallach2002}. One may also consider the mean linear entropy of two-qubit reduced density matrices, known as generalized global entanglement (GGE). Both quantities have been studied for the toric code in the presence of a nonlinear perturbation. They exhibit similar behavior such that their values remain close to unity in the topological phase and then drop to zero at the critical point. However, the difference between GE and GGE defines the conditional global entanglement, which exhibits a peak at the transition point~\cite{samimi2022global}.
	
To compute the global entanglement, we must calculate the linear entropy for the density matrices corresponding to different qubits at the output of our variational quantum circuit. Here, we note that the linear entropy is an approximation of the von Neumann entropy, $S(\rho) = -\operatorname{tr}(\rho \log \rho)$, and therefore, we can compute the mean value of the von Neumann entropy on the single-qubit density matrices. Evaluating the von Neumann entropy of a given density matrix typically requires diagonalizing it. However, in our setup, the state vector corresponding to the ground state is not directly accessible. Instead, we only have a quantum circuit whose output is the ground state of the system, meaning all information about this state must be derived by performing suitable measurements. This raises the question of how the von Neumann entropy can be determined via a set of measurements on the output of the circuit. In the following, we show that a second variational circuit can be employed to compute the global entanglement.

	We employ a variational method to estimate the von Neumann entropy based on measurement statistics. This approach relies on the well-known inequality for the entropy change under measurement,
	\begin{equation}
		S(\rho) \leq H(\{p_i\}) + \sum_i p_i S(\rho_i),
	\end{equation}
	where $p_i$ denotes the probability of obtaining outcome $i$, $\rho_i$ is the conditional state after observing that outcome, and $H(\{p_i\})$ is the Shannon entropy of the measurement outcomes~\cite{NielsenChuang}.
	
For a complete projective measurement in an orthonormal basis with projectors $P_i = |i\rangle\langle i|$, the post-measurement state corresponding to outcome $i$ is given by $\rho_i = |i\rangle\langle i|$, which is a pure state. Consequently, $S(\rho_i)=0$, and the above inequality reduces to $S(\rho) \leq H(\{p_i\})$. Thus, the Shannon entropy obtained from measurements in an arbitrary basis provides an upper bound on the von Neumann entropy. When the measurement basis coincides with the eigenbasis of $\rho$, the equality
\begin{equation}
	\label{fh}
	S(\rho) = \min_{B} H(\{p_i\})
\end{equation}
is attained. This observation motivates a simple variational approach for computing the von Neumann entropy. We first prepare the ground state of the system using the optimized parameters obtained from the VQE procedure, denoted by $\theta^*$, which generates the state $|\psi(\theta^*;h)\rangle$. The reduced state associated with each qubit is therefore a single-qubit density matrix. To compute the von Neumann entropy of such a state, we apply a single-qubit rotation $U(\alpha)$ to the corresponding qubit. A measurement in the $Z$ basis after this rotation is equivalent to a measurement in a rotated basis. From the measurement statistics, we then obtain the probability distribution $P(\alpha)=\{P_0(\alpha),P_1(\alpha)\}$. By computing the Shannon entropy of this distribution and varying the parameter $\alpha$, its minimum value can be determined, which provides an estimate of the von Neumann entropy of the subsystem according to Eq.~\eqref{fh}. Then, we derive mean value of this quantity for different qubits to compute the global entanglement.

\begin{figure*}[t]
	\centering
	\begin{subfigure}[t]{0.32\textwidth}
		\centering
		\includegraphics[width=\linewidth]{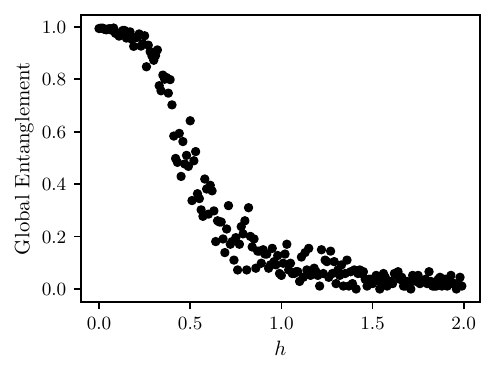}
		\caption{}
		\label{fig:8q_entropy}
	\end{subfigure}
	\hfill
	\begin{subfigure}[t]{0.32\textwidth}
		\centering
		\includegraphics[width=\linewidth]{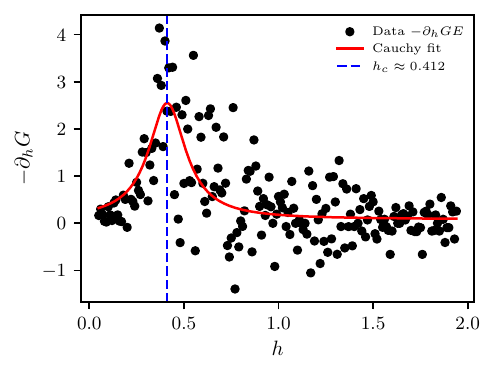}
		\caption{}
		\label{fig:8q_entropy_derivative_fit}
	\end{subfigure}
	\hfill
	\begin{subfigure}[t]{0.32\textwidth}
		\centering
		\includegraphics[width=\linewidth]{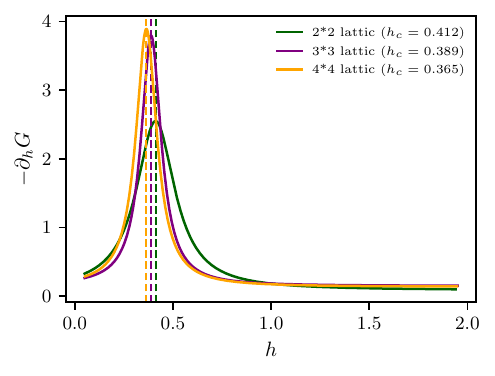}
		\caption{}
		\label{fig:8_18_24q_second_derivative_entropy}
	\end{subfigure}
	
\caption{Analysis of the global entanglement for different lattice sizes.
	(a) Global entanglement $G(h)$ as a function of the magnetic field $h$ for the $2\times2$ lattice.
	(b) Numerical derivative $-dG/dh$ for the $2\times2$ lattice, smoothed using a five-point moving average, where the peak indicates the maximum structural change.
	(c) Negative derivative $-dG/dh$ obtained from fitted entropy curves for the $2\times2$, $3\times3$, and $4\times4$ lattices, highlighting the phase-transition region. For all lattice sizes, 1024 measurement shots were employed for the single-qubit projective measurements in the variational entropy minimization procedure.}
	\label{fig:combined_entropy_analysis}
\end{figure*}

The behavior of the global entanglement as a function of the magnetic field $h$ for the $2\times2$ lattice is shown in Fig.~\ref{fig:8q_entropy}. In the weak-field regime ($h \approx 0$), the GE is close to $1$, which is the maximum possible value for a single qubit when the logarithm is taken in base $2$. This indicates that the qubit is nearly maximally entangled with the rest of the system in the ground state. As the field strength increases, the GE decreases continuously and approaches zero at larger values of $h$. This monotonic decrease reflects the crossover of the system from the topologically entangled regime to a polarized regime, in which the qubits progressively align with the external magnetic field and the entanglement is suppressed.

To estimate the location of this change of regime, we computed the numerical derivative of GE, $-dG/dh$. To reduce fluctuations in the numerical data, a five-point moving-average smoothing procedure was applied. The resulting $-dG/dh$ curve exhibits a pronounced peak, corresponding to the maximum rate of change of GE with respect to the magnetic field, as shown in Fig.~\ref{fig:8q_entropy_derivative_fit}. Such a peak signals the region in which the ground state undergoes its most rapid structural change under variation of the control parameter $h$.

To determine the peak position quantitatively, the derivative data were fitted with a Cauchy function, which provides a robust estimate of the peak center due to its symmetric profile around the maximum. The center of the fit, denoted by $h_c^{(G)}$, is in good quantitative agreement with the value extracted from the derivative of the ground-state energy for the same lattice, $h_c^{(E)}$. This consistency indicates that GE captures the same characteristic field scale identified by energetic observables.

Although the $2\times2$ lattice does not exhibit a true phase transition, the observed peak in $-dG/dh$ can be interpreted as a finite-size precursor of the topological-to-polarized transition expected in the thermodynamic limit. The agreement between $h_c^{(G)}$ and $h_c^{(E)}$ further supports the use of GE as an information-theoretic indicator of the transition region in finite systems.
	
To investigate finite-size effects, the calculations were repeated for several lattice sizes, $L \times L$. As $L$ increases, the peak associated with $-dG/dh$ becomes progressively more pronounced, and its position, denoted by $h_c(L)$, systematically shifts toward smaller magnetic-field values, as shown in Fig.~\ref{fig:8_18_24q_second_derivative_entropy}.

\begin{figure}[h]
	\centering
	\includegraphics[width=\linewidth]{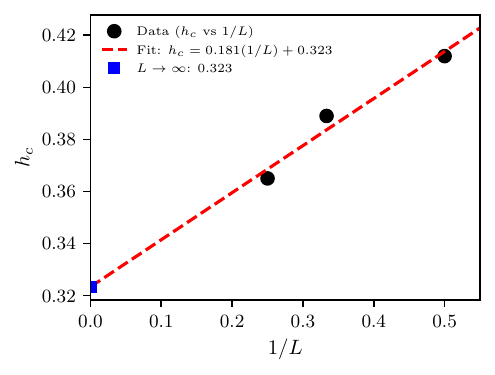}
	\caption{Finite-size scaling analysis of the critical magnetic field $h_c$, derived from the entropy data, as a function of the inverse lattice size $1/L$. Linear extrapolation (dashed line) to $1/L \to 0$ yields $h_c(\infty) \approx 0.323$ (blue square).}
	\label{fig:finite_size_scaling_entropy}
\end{figure}

To extrapolate the critical field in the thermodynamic limit ($L \rightarrow \infty$), the values of $h_c(L)$ were plotted as a function of $1/L$, as shown in Fig.~\ref{fig:finite_size_scaling_entropy}. A linear fit to these data yields $h_c(\infty)=0.323$, which is consistent with the result previously obtained from the analysis of the ground-state energy. This agreement indicates that the entanglement-based indicator extracted from the variational circuit successfully captures the scaling behavior of the critical field and provides a reliable estimate of $h_c$ in the thermodynamic limit.

We also examine the conditional global entanglement, defined as the difference between the generalized global entanglement and the global entanglement,
\begin{equation}
	\Delta G =
	\left|
	GGE-GE
	\right|.
\end{equation}

	Previous studies have shown that this quantity can exhibit a pronounced maximum in the vicinity of a phase transition~\cite{samimi2022global}. Here, we employ $\Delta G$ to track changes in the entanglement structure of the system as a function of the magnetic field.
	
\begin{figure}[h]
		\centering
		\includegraphics[width=8.7cm,height=2cm,angle=0]{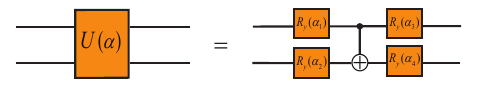}
		\caption{Parameterized two-qubit unitary $U_{AB}(\alpha_1,\alpha_2,\alpha_3,\alpha_4)$ employed in the entropy-estimation procedure. The circuit is composed of a CNOT gate together with four independent single-qubit rotations in a sense that $R_y(\alpha_1)$ and $R_y(\alpha_2)$ applied before the entangling gate, and $R_y(\alpha_3)$ and $R_y(\alpha_4)$ applied afterward.}
		\label{fig:two_qubit_entropy_circuit}
	\end{figure}

	To compute GGE we should quantify mean value of the normalized entropy of all two-qubit density matrices of the system. To compute the entropy of a two-qubit density matrix of the system, the variational approach introduced for the single-qubit subsystem was generalized to a two-qubit subsystem $AB$. Specifically, a parameterized unitary transformation $U_{AB}(\boldsymbol{\alpha})$ was applied exclusively to the selected qubit pair (see Fig.~\ref{fig:two_qubit_entropy_circuit}), generating a probability distribution $P(\alpha)=\{p_{00},p_{01},p_{10},p_{11}\}$, from which the corresponding Shannon entropy was computed.

\begin{figure}[h]
		\centering
		\includegraphics[width=\linewidth]{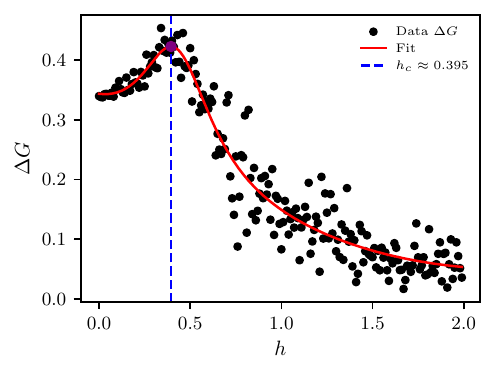}
		\caption{Conditional global entanglement $\Delta G = |GGE-GE|$ as a function of the magnetic field $h$ for the $2\times 2$ lattice. The black circles represent the raw data, and the solid red curve shows the numerical fit. The purple circle marks the estimated critical field at $h_c \approx 0.395$, corresponding to the maximum of the fitted curve.}
		\label{fig:8q_delta_s}
	\end{figure}

	As shown in Fig.~\ref{fig:8q_delta_s}, $\Delta G$ exhibits a clear maximum as a function of the magnetic field $h$, located near the transition region. The emergence of this peak indicates that, in the critical regime, the distinction between single- and two-qubit entanglement structures becomes more pronounced. This behavior is consistent with the enhancement of critical correlations in the vicinity of the phase transition.
	
	A numerical fit of the $\Delta G$ data using an appropriate peak-fitting model yields a peak position at $h_c \approx 0.395$. This value is in good agreement with the critical field extracted from the energy derivative and from the single-qubit entropy analysis. The consistency among these independent indicators suggests that the differential quantity $\Delta G$ serves as a sensitive probe of critical behavior within the VQE framework.
	
	\section{Robustness of the Topological Phase under Hardware Noise}\label{sec5}
	
	In the previous sections, the emergence of topological order was analyzed under ideal conditions using simulations without hardware noise. In practice, however, the realization of such phases on current noisy intermediate-scale quantum (NISQ) processors is inherently limited by hardware imperfections. In addition to statistical uncertainties arising from finite sampling—due to the probabilistic nature of quantum measurements and the limited number of shots—real devices are affected by decoherence and systematic operational errors during gate execution. In this section, we investigate the robustness of the topological phase transition under a realistic hardware noise model in order to estimate the shift of the critical field $h_c$ and the stability of the phase in the presence of these errors.
	
To capture the dominant imperfections of contemporary quantum hardware, we incorporate a composite noise model including both gate and readout errors. Single-qubit and two-qubit gate imperfections are modeled by depolarizing channels with separate error rates. For a single-qubit gate, the noisy operation is described by
\begin{equation}
	\mathcal{E}_{1}(\rho) = (1-p_1)\rho + \frac{p_1}{3}\sum_{i\in\{X,Y,Z\}}\sigma_i \rho \sigma_i,
\end{equation}
where $p_1$ is the single-qubit error probability. In addition, each CNOT gate is followed by a standard two-qubit depolarizing channel with error probability $p_2$. In the baseline noise model, we choose $p_1=0.005$ and $p_2=0.01$, consistent with typical error magnitudes reported for IBM Quantum devices.

Measurement errors are described by a symmetric readout channel,
\begin{equation}
	M=
	\begin{pmatrix}
		1-p_{\mathrm{read}} & p_{\mathrm{read}} \\
		p_{\mathrm{read}} & 1-p_{\mathrm{read}}
	\end{pmatrix},
\end{equation}
with $p_{\mathrm{read}}=0.01$. To examine the robustness of the results against increasing hardware noise, we introduce a strength factor $s$ and rescale all error probabilities as
\begin{equation}
	\{p_1,p_2,p_{\mathrm{read}}\} \rightarrow s \times \{0.005,0.01,0.01\},
\end{equation}
with $s \in \{0.5,1,1.5,2,2.5,3\}$. This framework allows us to track the evolution of the topological phase boundary from the weak-noise regime to the noise-dominated regime.
	
\begin{figure}[h]
		\centering
		\includegraphics[width=\linewidth]{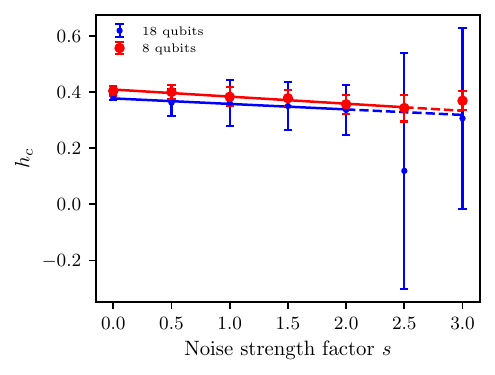}
		\caption{Critical magnetic field $h_c$ versus noise strength factor $s$ for $2\times2$ (8-qubit) and $3\times3$ (18-qubit) lattices, obtained using 1024 and 64 measurement shots, respectively. Solid lines indicate linear fits within the reliable noise regime, while dashed lines show the extrapolation. Error bars denote statistical uncertainty.}
		\label{fig:noise_scaling}
	\end{figure}

	Figure~\ref{fig:noise_scaling} shows the dependence of the critical field $h_c$ on the noise strength factor $s$ for the $2\times2$ ($8$-qubit) and $3\times3$ ($18$-qubit) lattices. At higher noise levels—specifically $s > 2.5$ for the $2\times2$ lattice and $s > 2$ for the $3\times3$ lattice—the data points deviate from the approximately linear trend. This deviation reflects the reduction of the signal-to-noise ratio, where statistical sampling uncertainties begin to dominate over the physical signal associated with the phase transition. These points are therefore excluded from the final fits.

	Within the low noise range, the critical field decreases linearly with increasing noise. This linear behavior indicates that topological phase displays a notable robustness against noise. In particular, a linear extrapolation within the reliable noise region, as $h_c(s) \approx h_c(0) - \kappa s$, allows us to estimate the value of $s$ for which $h_c \rightarrow 0$, where the observable signature of the topological phase effectively disappears. Although potential nonlinear behaviors at higher noise levels prevent this linear extrapolation from capturing the exact physical breakdown of the phase transition, it might nonetheless be regarded as a rough, operational upper bound for the circuit's noise resilience. For the $2 \times 2$ ($8$-qubit) system, the extrapolation yields $s \approx 16.2$, which corresponds to an effective error rate of roughly $\sim 0.16$. Despite being merely a heuristic estimate, this value reasonably aligns with the known toric code threshold ($\sim 0.1$), suggesting that beyond such noise levels, hardware errors would likely obscure any observable topological signatures.

Due to the high computational overhead of noisy simulations in the larger Hilbert space, the number of measurement shots for the $18$-qubit system was reduced to $64$, which naturally elevates the statistical uncertainty associated with shot noise. Nevertheless, the uncertainty growth observed in the $18$-qubit lattice is markedly more severe compared to the $8$-qubit case (especially for $s \ge 2.5$). This distinct behavior indicates that, beyond finite-sampling limitations, the pronounced error bars primarily reflect the larger circuit depth and the extensive accumulation of gate errors at higher noise levels. Overall, this analysis provides an indicative reference for the behavior of small-scale topological states under noise, contributing to a better understanding of circuit depth constraints in noisy quantum hardware.

\section{Conclusion}\label{sec6}
The efficacy of variational quantum algorithms in many-body physics crucially depends on designing a faithful trial ansatz. In particular, it is essential that the main physical properties of the system are well encoded by the ansatz. This becomes particularly challenging on near-term quantum hardware, where optimization must rely on a finite budget of measurement shots. Here, we showed that a suitable variational quantum circuit is able to simulate the ground state of the toric code model in the presence of a magnetic field, such that the main topological properties of the system are preserved. In particular, we demonstrated that the topological phase transition in this model is well detected by finite-size scaling analysis of the ground-state energy, evaluated with a finite number of measurement shots. We also introduced a simple variational circuit for the shot-based calculation of global entanglement in the ground state and showed that the topological nature of the phase transition is clearly identified by this entanglement measure. Finally, we considered the effect of noise on our analysis to reflect realistic experimental conditions. We showed that the transition point decreases linearly with increasing noise strength, serving as a signature of the robustness of the topological phase against noise. Our results reveal that by designing suitable parameterized quantum circuits for various quantum many-body systems, near-term quantum computers can be effectively utilized to probe important properties such as the entanglement structure in complex quantum states.\\

	\section*{Acknowledgement} 
	 We thank A. Ramezanpour for his useful comments on the paper.
	 
	 \section*{Funding}
	 This work is based upon research funded by Iran National Science Foundation (INSF) under project No 40403155.

\bibliography{refs}

\begin{thebibliography}{73}%
\makeatletter
\providecommand \@ifxundefined [1]{%
 \@ifx{#1\undefined}
}%
\providecommand \@ifnum [1]{%
 \ifnum #1\expandafter \@firstoftwo
 \else \expandafter \@secondoftwo
 \fi
}%
\providecommand \@ifx [1]{%
 \ifx #1\expandafter \@firstoftwo
 \else \expandafter \@secondoftwo
 \fi
}%
\providecommand \natexlab [1]{#1}%
\providecommand \enquote  [1]{``#1''}%
\providecommand \bibnamefont  [1]{#1}%
\providecommand \bibfnamefont [1]{#1}%
\providecommand \citenamefont [1]{#1}%
\providecommand \href@noop [0]{\@secondoftwo}%
\providecommand \href [0]{\begingroup \@sanitize@url \@href}%
\providecommand \@href[1]{\@@startlink{#1}\@@href}%
\providecommand \@@href[1]{\endgroup#1\@@endlink}%
\providecommand \@sanitize@url [0]{\catcode `\\12\catcode `\$12\catcode
  `\&12\catcode `\#12\catcode `\^12\catcode `\_12\catcode `\%12\relax}%
\providecommand \@@startlink[1]{}%
\providecommand \@@endlink[0]{}%
\providecommand \url  [0]{\begingroup\@sanitize@url \@url }%
\providecommand \@url [1]{\endgroup\@href {#1}{\urlprefix }}%
\providecommand \urlprefix  [0]{URL }%
\providecommand \Eprint [0]{\href }%
\providecommand \doibase [0]{https://doi.org/}%
\providecommand \selectlanguage [0]{\@gobble}%
\providecommand \bibinfo  [0]{\@secondoftwo}%
\providecommand \bibfield  [0]{\@secondoftwo}%
\providecommand \translation [1]{[#1]}%
\providecommand \BibitemOpen [0]{}%
\providecommand \bibitemStop [0]{}%
\providecommand \bibitemNoStop [0]{.\EOS\space}%
\providecommand \EOS [0]{\spacefactor3000\relax}%
\providecommand \BibitemShut  [1]{\csname bibitem#1\endcsname}%
\let\auto@bib@innerbib\@empty
\bibitem [{\citenamefont {Sachdev}(2023)}]{sachdev2023quantum}%
  \BibitemOpen
  \bibfield  {author} {\bibinfo {author} {\bibfnamefont {S.}~\bibnamefont
  {Sachdev}},\ }\href {https://doi.org/10.1017/9781009212717} {\emph {\bibinfo
  {title} {Quantum Phases of Matter}}}\ (\bibinfo  {publisher} {Cambridge
  University Press},\ \bibinfo {address} {Cambridge},\ \bibinfo {year}
  {2023})\BibitemShut {NoStop}%
\bibitem [{\citenamefont {Wen}(1990)}]{wen1990topological}%
  \BibitemOpen
  \bibfield  {author} {\bibinfo {author} {\bibfnamefont {X.-G.}\ \bibnamefont
  {Wen}},\ }\bibfield  {title} {\bibinfo {title} {Topological orders in rigid
  states},\ }\href {https://doi.org/10.1142/S0217979290000139} {\bibfield
  {journal} {\bibinfo  {journal} {Int. J. Mod. Phys. B}\ }\textbf {\bibinfo
  {volume} {4}},\ \bibinfo {pages} {239} (\bibinfo {year} {1990})}\BibitemShut
  {NoStop}%
\bibitem [{\citenamefont {Wen}(2004)}]{Wen1}%
  \BibitemOpen
  \bibfield  {author} {\bibinfo {author} {\bibfnamefont {X.-G.}\ \bibnamefont
  {Wen}},\ }\href {https://doi.org/10.1093/acprof:oso/9780199227259.001.0001}
  {\emph {\bibinfo {title} {Quantum Field Theory of Many-Body Systems: From the
  Origin of Sound to an Origin of Light and Electrons}}}\ (\bibinfo
  {publisher} {Oxford University Press},\ \bibinfo {address} {Oxford},\
  \bibinfo {year} {2004})\BibitemShut {NoStop}%
\bibitem [{\citenamefont {Wen}(2017)}]{Wen2}%
  \BibitemOpen
  \bibfield  {author} {\bibinfo {author} {\bibfnamefont {X.-G.}\ \bibnamefont
  {Wen}},\ }\bibfield  {title} {\bibinfo {title} {Colloquium: {Zoo} of
  quantum-topological phases of matter},\ }\href
  {https://doi.org/10.1103/RevModPhys.89.041004} {\bibfield  {journal}
  {\bibinfo  {journal} {Rev. Mod. Phys.}\ }\textbf {\bibinfo {volume} {89}},\
  \bibinfo {pages} {041004} (\bibinfo {year} {2017})}\BibitemShut {NoStop}%
\bibitem [{\citenamefont {Levin}\ and\ \citenamefont
  {Wen}(2006)}]{levin2006detecting}%
  \BibitemOpen
  \bibfield  {author} {\bibinfo {author} {\bibfnamefont {M.}~\bibnamefont
  {Levin}}\ and\ \bibinfo {author} {\bibfnamefont {X.-G.}\ \bibnamefont
  {Wen}},\ }\bibfield  {title} {\bibinfo {title} {Detecting topological order
  in a ground state wave function},\ }\href
  {https://doi.org/10.1103/PhysRevLett.96.110405} {\bibfield  {journal}
  {\bibinfo  {journal} {Phys. Rev. Lett.}\ }\textbf {\bibinfo {volume} {96}},\
  \bibinfo {pages} {110405} (\bibinfo {year} {2006})}\BibitemShut {NoStop}%
\bibitem [{\citenamefont {Kitaev}(2003)}]{kitaev2003}%
  \BibitemOpen
  \bibfield  {author} {\bibinfo {author} {\bibfnamefont {A.~Y.}\ \bibnamefont
  {Kitaev}},\ }\bibfield  {title} {\bibinfo {title} {Fault-tolerant quantum
  computation by anyons},\ }\href
  {https://doi.org/10.1016/S0003-4916(02)00018-0} {\bibfield  {journal}
  {\bibinfo  {journal} {Ann. Phys.}\ }\textbf {\bibinfo {volume} {303}},\
  \bibinfo {pages} {2} (\bibinfo {year} {2003})}\BibitemShut {NoStop}%
\bibitem [{\citenamefont {Troyer}\ and\ \citenamefont
  {Wiese}(2005)}]{troyer2005computational}%
  \BibitemOpen
  \bibfield  {author} {\bibinfo {author} {\bibfnamefont {M.}~\bibnamefont
  {Troyer}}\ and\ \bibinfo {author} {\bibfnamefont {U.-J.}\ \bibnamefont
  {Wiese}},\ }\bibfield  {title} {\bibinfo {title} {Computational complexity
  and fundamental limitations to fermionic quantum {Monte Carlo} simulations},\
  }\href {https://doi.org/10.1103/PhysRevLett.94.170201} {\bibfield  {journal}
  {\bibinfo  {journal} {Phys. Rev. Lett.}\ }\textbf {\bibinfo {volume} {94}},\
  \bibinfo {pages} {170201} (\bibinfo {year} {2005})}\BibitemShut {NoStop}%
\bibitem [{\citenamefont {Park}\ \emph {et~al.}(2016)\citenamefont {Park},
  \citenamefont {McKay}, \citenamefont {Lu},\ and\ \citenamefont
  {Laflamme}}]{park2016simulation}%
  \BibitemOpen
  \bibfield  {author} {\bibinfo {author} {\bibfnamefont {A.~J.}\ \bibnamefont
  {Park}}, \bibinfo {author} {\bibfnamefont {E.}~\bibnamefont {McKay}},
  \bibinfo {author} {\bibfnamefont {D.}~\bibnamefont {Lu}},\ and\ \bibinfo
  {author} {\bibfnamefont {R.}~\bibnamefont {Laflamme}},\ }\bibfield  {title}
  {\bibinfo {title} {Simulation of anyonic statistics and its topological path
  independence using a seven-qubit quantum simulator},\ }\href
  {https://doi.org/10.1088/1367-2630/18/4/043043} {\bibfield  {journal}
  {\bibinfo  {journal} {New J. Phys.}\ }\textbf {\bibinfo {volume} {18}},\
  \bibinfo {pages} {043043} (\bibinfo {year} {2016})}\BibitemShut {NoStop}%
\bibitem [{\citenamefont {Foss-Feig}\ \emph {et~al.}(2023)\citenamefont
  {Foss-Feig}, \citenamefont {Tikku}, \citenamefont {Lu}, \citenamefont {Mayer}
  \emph {et~al.}}]{fossfeig2023adaptive}%
  \BibitemOpen
  \bibfield  {author} {\bibinfo {author} {\bibfnamefont {M.}~\bibnamefont
  {Foss-Feig}}, \bibinfo {author} {\bibfnamefont {A.}~\bibnamefont {Tikku}},
  \bibinfo {author} {\bibfnamefont {T.-C.}\ \bibnamefont {Lu}}, \bibinfo
  {author} {\bibfnamefont {K.}~\bibnamefont {Mayer}}, \emph {et~al.},\
  }\bibfield  {title} {\bibinfo {title} {Experimental demonstration of the
  advantage of adaptive quantum circuits},\ }\bibfield  {journal} {\bibinfo
  {journal} {arXiv:2302.03029}\ }\href
  {https://doi.org/10.48550/arXiv.2302.03029} {10.48550/arXiv.2302.03029}
  (\bibinfo {year} {2023})\BibitemShut {NoStop}%
\bibitem [{\citenamefont {Xu}\ \emph {et~al.}(2023)\citenamefont {Xu},
  \citenamefont {Sun}, \citenamefont {Wang}, \citenamefont {Xiang},
  \citenamefont {Bao}, \citenamefont {Zhu}, \citenamefont {Shen}, \citenamefont
  {Song}, \citenamefont {Gao}, \citenamefont {Wang} \emph
  {et~al.}}]{xu2024digital}%
  \BibitemOpen
  \bibfield  {author} {\bibinfo {author} {\bibfnamefont {S.}~\bibnamefont
  {Xu}}, \bibinfo {author} {\bibfnamefont {Z.-Z.}\ \bibnamefont {Sun}},
  \bibinfo {author} {\bibfnamefont {K.}~\bibnamefont {Wang}}, \bibinfo {author}
  {\bibfnamefont {L.}~\bibnamefont {Xiang}}, \bibinfo {author} {\bibfnamefont
  {Z.}~\bibnamefont {Bao}}, \bibinfo {author} {\bibfnamefont {Z.}~\bibnamefont
  {Zhu}}, \bibinfo {author} {\bibfnamefont {F.}~\bibnamefont {Shen}}, \bibinfo
  {author} {\bibfnamefont {Z.}~\bibnamefont {Song}}, \bibinfo {author}
  {\bibfnamefont {P.}~\bibnamefont {Gao}}, \bibinfo {author} {\bibfnamefont
  {C.}~\bibnamefont {Wang}}, \emph {et~al.},\ }\bibfield  {title} {\bibinfo
  {title} {Digital simulation of projective non-{Abelian} anyons with 68
  superconducting qubits},\ }\href
  {https://doi.org/10.1088/0256-307X/40/6/060301} {\bibfield  {journal}
  {\bibinfo  {journal} {Chin. Phys. Lett.}\ }\textbf {\bibinfo {volume} {40}},\
  \bibinfo {pages} {060301} (\bibinfo {year} {2023})}\BibitemShut {NoStop}%
\bibitem [{\citenamefont {Semeghini}\ \emph {et~al.}(2021)\citenamefont
  {Semeghini}, \citenamefont {Levine}, \citenamefont {Keesling}, \citenamefont
  {Ebadi}, \citenamefont {Wang}, \citenamefont {Bluvstein}, \citenamefont
  {Verresen}, \citenamefont {Pichler}, \citenamefont {Kalinowski},
  \citenamefont {Zhou} \emph {et~al.}}]{semeghini2021probing}%
  \BibitemOpen
  \bibfield  {author} {\bibinfo {author} {\bibfnamefont {G.}~\bibnamefont
  {Semeghini}}, \bibinfo {author} {\bibfnamefont {H.}~\bibnamefont {Levine}},
  \bibinfo {author} {\bibfnamefont {A.}~\bibnamefont {Keesling}}, \bibinfo
  {author} {\bibfnamefont {S.}~\bibnamefont {Ebadi}}, \bibinfo {author}
  {\bibfnamefont {T.~T.}\ \bibnamefont {Wang}}, \bibinfo {author}
  {\bibfnamefont {D.}~\bibnamefont {Bluvstein}}, \bibinfo {author}
  {\bibfnamefont {R.}~\bibnamefont {Verresen}}, \bibinfo {author}
  {\bibfnamefont {H.}~\bibnamefont {Pichler}}, \bibinfo {author} {\bibfnamefont
  {M.}~\bibnamefont {Kalinowski}}, \bibinfo {author} {\bibfnamefont
  {R.}~\bibnamefont {Zhou}}, \emph {et~al.},\ }\bibfield  {title} {\bibinfo
  {title} {Probing topological spin liquids on a programmable quantum
  simulator},\ }\href {https://doi.org/10.1126/science.abi8794} {\bibfield
  {journal} {\bibinfo  {journal} {Science}\ }\textbf {\bibinfo {volume}
  {374}},\ \bibinfo {pages} {1242} (\bibinfo {year} {2021})}\BibitemShut
  {NoStop}%
\bibitem [{\citenamefont {Barreiro}\ \emph {et~al.}(2011)\citenamefont
  {Barreiro}, \citenamefont {M{\"u}ller}, \citenamefont {Schindler},
  \citenamefont {Nigg}, \citenamefont {Monz}, \citenamefont {Chwalla},
  \citenamefont {Hennrich}, \citenamefont {Roos}, \citenamefont {Zoller},\ and\
  \citenamefont {Blatt}}]{barreiro2011open}%
  \BibitemOpen
  \bibfield  {author} {\bibinfo {author} {\bibfnamefont {J.~T.}\ \bibnamefont
  {Barreiro}}, \bibinfo {author} {\bibfnamefont {M.}~\bibnamefont
  {M{\"u}ller}}, \bibinfo {author} {\bibfnamefont {P.}~\bibnamefont
  {Schindler}}, \bibinfo {author} {\bibfnamefont {D.}~\bibnamefont {Nigg}},
  \bibinfo {author} {\bibfnamefont {T.}~\bibnamefont {Monz}}, \bibinfo {author}
  {\bibfnamefont {M.}~\bibnamefont {Chwalla}}, \bibinfo {author} {\bibfnamefont
  {M.}~\bibnamefont {Hennrich}}, \bibinfo {author} {\bibfnamefont {C.~F.}\
  \bibnamefont {Roos}}, \bibinfo {author} {\bibfnamefont {P.}~\bibnamefont
  {Zoller}},\ and\ \bibinfo {author} {\bibfnamefont {R.}~\bibnamefont
  {Blatt}},\ }\bibfield  {title} {\bibinfo {title} {An open-system quantum
  simulator with trapped ions},\ }\href {https://doi.org/10.1038/nature09801}
  {\bibfield  {journal} {\bibinfo  {journal} {Nature (London)}\ }\textbf
  {\bibinfo {volume} {470}},\ \bibinfo {pages} {486} (\bibinfo {year}
  {2011})}\BibitemShut {NoStop}%
\bibitem [{\citenamefont {Arute}\ \emph {et~al.}(2019)\citenamefont {Arute}
  \emph {et~al.}}]{arute2019quantum}%
  \BibitemOpen
  \bibfield  {author} {\bibinfo {author} {\bibfnamefont {F.}~\bibnamefont
  {Arute}} \emph {et~al.},\ }\bibfield  {title} {\bibinfo {title} {Quantum
  supremacy using a programmable superconducting processor},\ }\href
  {https://doi.org/10.1038/s41586-019-1666-5} {\bibfield  {journal} {\bibinfo
  {journal} {Nature}\ }\textbf {\bibinfo {volume} {574}},\ \bibinfo {pages}
  {505} (\bibinfo {year} {2019})}\BibitemShut {NoStop}%
\bibitem [{\citenamefont {Ebadi}\ \emph {et~al.}(2021)\citenamefont {Ebadi}
  \emph {et~al.}}]{ebadi2021quantum}%
  \BibitemOpen
  \bibfield  {author} {\bibinfo {author} {\bibfnamefont {S.}~\bibnamefont
  {Ebadi}} \emph {et~al.},\ }\bibfield  {title} {\bibinfo {title} {Quantum
  phases of matter on a 256-atom programmable quantum simulator},\ }\href
  {https://doi.org/10.1038/s41586-021-03582-4} {\bibfield  {journal} {\bibinfo
  {journal} {Nature}\ }\textbf {\bibinfo {volume} {595}},\ \bibinfo {pages}
  {227} (\bibinfo {year} {2021})}\BibitemShut {NoStop}%
\bibitem [{\citenamefont {Browaeys}\ and\ \citenamefont
  {Lahaye}(2020)}]{browaeys2020manybody}%
  \BibitemOpen
  \bibfield  {author} {\bibinfo {author} {\bibfnamefont {A.}~\bibnamefont
  {Browaeys}}\ and\ \bibinfo {author} {\bibfnamefont {T.}~\bibnamefont
  {Lahaye}},\ }\bibfield  {title} {\bibinfo {title} {Many-body physics with
  individually controlled {Rydberg} atoms},\ }\href
  {https://doi.org/10.1038/s41567-019-0733-z} {\bibfield  {journal} {\bibinfo
  {journal} {Nat. Phys.}\ }\textbf {\bibinfo {volume} {16}},\ \bibinfo {pages}
  {132} (\bibinfo {year} {2020})}\BibitemShut {NoStop}%
\bibitem [{\citenamefont {Altman}\ \emph {et~al.}(2021)\citenamefont {Altman},
  \citenamefont {Brown}, \citenamefont {Carleo}, \citenamefont {Carr},
  \citenamefont {Demler}, \citenamefont {Chin}, \citenamefont {DeMarco},
  \citenamefont {Economou}, \citenamefont {Fellous-Asiani}, \citenamefont
  {Greiner} \emph {et~al.}}]{altman2021quantum}%
  \BibitemOpen
  \bibfield  {author} {\bibinfo {author} {\bibfnamefont {E.}~\bibnamefont
  {Altman}}, \bibinfo {author} {\bibfnamefont {K.~R.}\ \bibnamefont {Brown}},
  \bibinfo {author} {\bibfnamefont {G.}~\bibnamefont {Carleo}}, \bibinfo
  {author} {\bibfnamefont {L.~D.}\ \bibnamefont {Carr}}, \bibinfo {author}
  {\bibfnamefont {E.}~\bibnamefont {Demler}}, \bibinfo {author} {\bibfnamefont
  {C.}~\bibnamefont {Chin}}, \bibinfo {author} {\bibfnamefont {B.}~\bibnamefont
  {DeMarco}}, \bibinfo {author} {\bibfnamefont {S.~E.}\ \bibnamefont
  {Economou}}, \bibinfo {author} {\bibfnamefont {D.}~\bibnamefont
  {Fellous-Asiani}}, \bibinfo {author} {\bibfnamefont {M.}~\bibnamefont
  {Greiner}}, \emph {et~al.},\ }\bibfield  {title} {\bibinfo {title} {Quantum
  simulators: {Architectures} and opportunities},\ }\href
  {https://doi.org/10.1103/PRXQuantum.2.017003} {\bibfield  {journal} {\bibinfo
   {journal} {PRX Quantum}\ }\textbf {\bibinfo {volume} {2}},\ \bibinfo {pages}
  {017003} (\bibinfo {year} {2021})}\BibitemShut {NoStop}%
\bibitem [{\citenamefont {Shi}\ \emph {et~al.}(2023)\citenamefont {Shi},
  \citenamefont {Liu}, \citenamefont {Zhang}, \citenamefont {Xiang},
  \citenamefont {Huang}, \citenamefont {Liu}, \citenamefont {Wang},
  \citenamefont {Zhang}, \citenamefont {Deng}, \citenamefont {Liang} \emph
  {et~al.}}]{shi2023quantum}%
  \BibitemOpen
  \bibfield  {author} {\bibinfo {author} {\bibfnamefont {Y.-H.}\ \bibnamefont
  {Shi}}, \bibinfo {author} {\bibfnamefont {Y.}~\bibnamefont {Liu}}, \bibinfo
  {author} {\bibfnamefont {Y.-R.}\ \bibnamefont {Zhang}}, \bibinfo {author}
  {\bibfnamefont {Z.}~\bibnamefont {Xiang}}, \bibinfo {author} {\bibfnamefont
  {K.}~\bibnamefont {Huang}}, \bibinfo {author} {\bibfnamefont
  {T.}~\bibnamefont {Liu}}, \bibinfo {author} {\bibfnamefont {Y.-Y.}\
  \bibnamefont {Wang}}, \bibinfo {author} {\bibfnamefont {J.-C.}\ \bibnamefont
  {Zhang}}, \bibinfo {author} {\bibfnamefont {C.-L.}\ \bibnamefont {Deng}},
  \bibinfo {author} {\bibfnamefont {G.-H.}\ \bibnamefont {Liang}}, \emph
  {et~al.},\ }\bibfield  {title} {\bibinfo {title} {Quantum simulation of
  topological zero modes on a 41-qubit superconducting processor},\ }\href
  {https://doi.org/10.1103/PhysRevLett.131.080401} {\bibfield  {journal}
  {\bibinfo  {journal} {Phys. Rev. Lett}\ }\textbf {\bibinfo {volume} {131}},\
  \bibinfo {pages} {080401} (\bibinfo {year} {2023})}\BibitemShut {NoStop}%
\bibitem [{\citenamefont {Peruzzo}\ \emph {et~al.}(2014)\citenamefont
  {Peruzzo}, \citenamefont {McClean}, \citenamefont {Shadbolt}, \citenamefont
  {Yung}, \citenamefont {Zhou}, \citenamefont {Love}, \citenamefont
  {Aspuru-Guzik},\ and\ \citenamefont {O'Brien}}]{peruzzo2014variational}%
  \BibitemOpen
  \bibfield  {author} {\bibinfo {author} {\bibfnamefont {A.}~\bibnamefont
  {Peruzzo}}, \bibinfo {author} {\bibfnamefont {J.}~\bibnamefont {McClean}},
  \bibinfo {author} {\bibfnamefont {P.}~\bibnamefont {Shadbolt}}, \bibinfo
  {author} {\bibfnamefont {M.-H.}\ \bibnamefont {Yung}}, \bibinfo {author}
  {\bibfnamefont {X.-Q.}\ \bibnamefont {Zhou}}, \bibinfo {author}
  {\bibfnamefont {P.~J.}\ \bibnamefont {Love}}, \bibinfo {author}
  {\bibfnamefont {A.}~\bibnamefont {Aspuru-Guzik}},\ and\ \bibinfo {author}
  {\bibfnamefont {J.~L.}\ \bibnamefont {O'Brien}},\ }\bibfield  {title}
  {\bibinfo {title} {A variational eigenvalue solver on a photonic quantum
  processor},\ }\href {https://doi.org/10.1038/ncomms5213} {\bibfield
  {journal} {\bibinfo  {journal} {Nat. Commun.}\ }\textbf {\bibinfo {volume}
  {5}},\ \bibinfo {pages} {4213} (\bibinfo {year} {2014})}\BibitemShut
  {NoStop}%
\bibitem [{\citenamefont {McClean}\ \emph {et~al.}(2016)\citenamefont
  {McClean}, \citenamefont {Romero}, \citenamefont {Babbush},\ and\
  \citenamefont {Aspuru-Guzik}}]{mcclean2016theory}%
  \BibitemOpen
  \bibfield  {author} {\bibinfo {author} {\bibfnamefont {J.~R.}\ \bibnamefont
  {McClean}}, \bibinfo {author} {\bibfnamefont {J.}~\bibnamefont {Romero}},
  \bibinfo {author} {\bibfnamefont {R.}~\bibnamefont {Babbush}},\ and\ \bibinfo
  {author} {\bibfnamefont {A.}~\bibnamefont {Aspuru-Guzik}},\ }\bibfield
  {title} {\bibinfo {title} {The theory of variational hybrid quantum-classical
  algorithms},\ }\href {https://doi.org/10.1088/1367-2630/18/2/023023}
  {\bibfield  {journal} {\bibinfo  {journal} {New J. Phys.}\ }\textbf {\bibinfo
  {volume} {18}},\ \bibinfo {pages} {023023} (\bibinfo {year}
  {2016})}\BibitemShut {NoStop}%
\bibitem [{\citenamefont {Zhang}\ \emph {et~al.}(2022)\citenamefont {Zhang},
  \citenamefont {Jiang}, \citenamefont {Deng}, \citenamefont {Wang},
  \citenamefont {Chen}, \citenamefont {Zhang}, \citenamefont {Ren},
  \citenamefont {Dong}, \citenamefont {Xu}, \citenamefont {Gao} \emph
  {et~al.}}]{zhang2022digital}%
  \BibitemOpen
  \bibfield  {author} {\bibinfo {author} {\bibfnamefont {X.}~\bibnamefont
  {Zhang}}, \bibinfo {author} {\bibfnamefont {W.}~\bibnamefont {Jiang}},
  \bibinfo {author} {\bibfnamefont {J.}~\bibnamefont {Deng}}, \bibinfo {author}
  {\bibfnamefont {K.}~\bibnamefont {Wang}}, \bibinfo {author} {\bibfnamefont
  {J.}~\bibnamefont {Chen}}, \bibinfo {author} {\bibfnamefont {P.}~\bibnamefont
  {Zhang}}, \bibinfo {author} {\bibfnamefont {W.}~\bibnamefont {Ren}}, \bibinfo
  {author} {\bibfnamefont {H.}~\bibnamefont {Dong}}, \bibinfo {author}
  {\bibfnamefont {S.}~\bibnamefont {Xu}}, \bibinfo {author} {\bibfnamefont
  {Y.}~\bibnamefont {Gao}}, \emph {et~al.},\ }\bibfield  {title} {\bibinfo
  {title} {Digital quantum simulation of floquet symmetry-protected topological
  phases},\ }\href {https://doi.org/10.1038/s41586-022-04854-3} {\bibfield
  {journal} {\bibinfo  {journal} {Nature}\ }\textbf {\bibinfo {volume} {607}},\
  \bibinfo {pages} {468} (\bibinfo {year} {2022})}\BibitemShut {NoStop}%
\bibitem [{\citenamefont {Zapletal}\ \emph {et~al.}(2024)\citenamefont
  {Zapletal}, \citenamefont {McMahon},\ and\ \citenamefont
  {Hartmann}}]{zapletal2024error}%
  \BibitemOpen
  \bibfield  {author} {\bibinfo {author} {\bibfnamefont {P.}~\bibnamefont
  {Zapletal}}, \bibinfo {author} {\bibfnamefont {N.~A.}\ \bibnamefont
  {McMahon}},\ and\ \bibinfo {author} {\bibfnamefont {M.~J.}\ \bibnamefont
  {Hartmann}},\ }\bibfield  {title} {\bibinfo {title} {Error-tolerant quantum
  convolutional neural networks for symmetry-protected topological phases},\
  }\href {https://doi.org/10.1103/PhysRevResearch.6.033111} {\bibfield
  {journal} {\bibinfo  {journal} {Phys. Rev. Res}\ }\textbf {\bibinfo {volume}
  {6}},\ \bibinfo {pages} {033111} (\bibinfo {year} {2024})}\BibitemShut
  {NoStop}%
\bibitem [{\citenamefont {Kandala}\ \emph {et~al.}(2017)\citenamefont
  {Kandala}, \citenamefont {Mezzacapo}, \citenamefont {Temme}, \citenamefont
  {Takita}, \citenamefont {Brink}, \citenamefont {Chow},\ and\ \citenamefont
  {Gambetta}}]{kandala2017hardware}%
  \BibitemOpen
  \bibfield  {author} {\bibinfo {author} {\bibfnamefont {A.}~\bibnamefont
  {Kandala}}, \bibinfo {author} {\bibfnamefont {A.}~\bibnamefont {Mezzacapo}},
  \bibinfo {author} {\bibfnamefont {K.}~\bibnamefont {Temme}}, \bibinfo
  {author} {\bibfnamefont {M.}~\bibnamefont {Takita}}, \bibinfo {author}
  {\bibfnamefont {M.}~\bibnamefont {Brink}}, \bibinfo {author} {\bibfnamefont
  {J.~M.}\ \bibnamefont {Chow}},\ and\ \bibinfo {author} {\bibfnamefont
  {J.~M.}\ \bibnamefont {Gambetta}},\ }\bibfield  {title} {\bibinfo {title}
  {Hardware-efficient variational quantum eigensolver for small molecules and
  quantum magnets},\ }\href {https://doi.org/10.1038/nature23879} {\bibfield
  {journal} {\bibinfo  {journal} {Nature}\ }\textbf {\bibinfo {volume} {549}},\
  \bibinfo {pages} {242} (\bibinfo {year} {2017})}\BibitemShut {NoStop}%
\bibitem [{\citenamefont {Nam}\ \emph {et~al.}(2020)\citenamefont {Nam},
  \citenamefont {Chen}, \citenamefont {Pisenti}, \citenamefont {Wright},
  \citenamefont {Delaney}, \citenamefont {Maslov}, \citenamefont {Brown},
  \citenamefont {Allen}, \citenamefont {Amini}, \citenamefont {Apisdorf} \emph
  {et~al.}}]{nam2020ground}%
  \BibitemOpen
  \bibfield  {author} {\bibinfo {author} {\bibfnamefont {Y.}~\bibnamefont
  {Nam}}, \bibinfo {author} {\bibfnamefont {J.-S.}\ \bibnamefont {Chen}},
  \bibinfo {author} {\bibfnamefont {N.~C.}\ \bibnamefont {Pisenti}}, \bibinfo
  {author} {\bibfnamefont {K.}~\bibnamefont {Wright}}, \bibinfo {author}
  {\bibfnamefont {C.}~\bibnamefont {Delaney}}, \bibinfo {author} {\bibfnamefont
  {D.}~\bibnamefont {Maslov}}, \bibinfo {author} {\bibfnamefont {K.~R.}\
  \bibnamefont {Brown}}, \bibinfo {author} {\bibfnamefont {S.}~\bibnamefont
  {Allen}}, \bibinfo {author} {\bibfnamefont {J.~M.}\ \bibnamefont {Amini}},
  \bibinfo {author} {\bibfnamefont {J.}~\bibnamefont {Apisdorf}}, \emph
  {et~al.},\ }\bibfield  {title} {\bibinfo {title} {Ground-state energy
  estimation of the water molecule on a trapped-ion quantum computer},\ }\href
  {https://doi.org/10.1038/s41534-020-0259-3} {\bibfield  {journal} {\bibinfo
  {journal} {npj Quantum Inf}\ }\textbf {\bibinfo {volume} {6}},\ \bibinfo
  {pages} {33} (\bibinfo {year} {2020})}\BibitemShut {NoStop}%
\bibitem [{\citenamefont {Cerezo}\ \emph {et~al.}(2021)\citenamefont {Cerezo},
  \citenamefont {Arrasmith}, \citenamefont {Babbush}, \citenamefont {Benjamin},
  \citenamefont {Endo}, \citenamefont {Fujii}, \citenamefont {McClean},
  \citenamefont {Mitarai}, \citenamefont {Yuan}, \citenamefont {Cincio},\ and\
  \citenamefont {Coles}}]{cerezo2021variational}%
  \BibitemOpen
  \bibfield  {author} {\bibinfo {author} {\bibfnamefont {M.}~\bibnamefont
  {Cerezo}}, \bibinfo {author} {\bibfnamefont {A.}~\bibnamefont {Arrasmith}},
  \bibinfo {author} {\bibfnamefont {R.}~\bibnamefont {Babbush}}, \bibinfo
  {author} {\bibfnamefont {S.~C.}\ \bibnamefont {Benjamin}}, \bibinfo {author}
  {\bibfnamefont {S.}~\bibnamefont {Endo}}, \bibinfo {author} {\bibfnamefont
  {K.}~\bibnamefont {Fujii}}, \bibinfo {author} {\bibfnamefont {J.~R.}\
  \bibnamefont {McClean}}, \bibinfo {author} {\bibfnamefont {K.}~\bibnamefont
  {Mitarai}}, \bibinfo {author} {\bibfnamefont {X.}~\bibnamefont {Yuan}},
  \bibinfo {author} {\bibfnamefont {L.}~\bibnamefont {Cincio}},\ and\ \bibinfo
  {author} {\bibfnamefont {P.~J.}\ \bibnamefont {Coles}},\ }\bibfield  {title}
  {\bibinfo {title} {Variational quantum algorithms},\ }\href
  {https://doi.org/10.1038/s42254-021-00348-9} {\bibfield  {journal} {\bibinfo
  {journal} {Nat. Rev. Phys.}\ }\textbf {\bibinfo {volume} {3}},\ \bibinfo
  {pages} {625} (\bibinfo {year} {2021})}\BibitemShut {NoStop}%
\bibitem [{\citenamefont {Yuan}\ \emph {et~al.}(2019)\citenamefont {Yuan},
  \citenamefont {Endo}, \citenamefont {Zhao}, \citenamefont {Li},\ and\
  \citenamefont {Benjamin}}]{yuan2019theory}%
  \BibitemOpen
  \bibfield  {author} {\bibinfo {author} {\bibfnamefont {X.}~\bibnamefont
  {Yuan}}, \bibinfo {author} {\bibfnamefont {S.}~\bibnamefont {Endo}}, \bibinfo
  {author} {\bibfnamefont {Q.}~\bibnamefont {Zhao}}, \bibinfo {author}
  {\bibfnamefont {Y.}~\bibnamefont {Li}},\ and\ \bibinfo {author}
  {\bibfnamefont {S.~C.}\ \bibnamefont {Benjamin}},\ }\bibfield  {title}
  {\bibinfo {title} {Theory of variational quantum simulation},\ }\href
  {https://doi.org/10.22331/q-2019-10-07-191} {\bibfield  {journal} {\bibinfo
  {journal} {Quantum}\ }\textbf {\bibinfo {volume} {3}},\ \bibinfo {pages}
  {191} (\bibinfo {year} {2019})}\BibitemShut {NoStop}%
\bibitem [{\citenamefont {Lyu}\ \emph {et~al.}(2023)\citenamefont {Lyu},
  \citenamefont {Xu}, \citenamefont {Yung},\ and\ \citenamefont
  {Bayat}}]{lyu2023symmetry}%
  \BibitemOpen
  \bibfield  {author} {\bibinfo {author} {\bibfnamefont {C.}~\bibnamefont
  {Lyu}}, \bibinfo {author} {\bibfnamefont {X.}~\bibnamefont {Xu}}, \bibinfo
  {author} {\bibfnamefont {M.-H.}\ \bibnamefont {Yung}},\ and\ \bibinfo
  {author} {\bibfnamefont {A.}~\bibnamefont {Bayat}},\ }\bibfield  {title}
  {\bibinfo {title} {Symmetry enhanced variational quantum spin eigensolver},\
  }\href {https://doi.org/10.22331/q-2023-01-19-899} {\bibfield  {journal}
  {\bibinfo  {journal} {Quantum}\ }\textbf {\bibinfo {volume} {7}},\ \bibinfo
  {pages} {899} (\bibinfo {year} {2023})}\BibitemShut {NoStop}%
\bibitem [{\citenamefont {Lyu}\ \emph {et~al.}(2020)\citenamefont {Lyu},
  \citenamefont {Montenegro},\ and\ \citenamefont
  {Bayat}}]{lyu2020accelerated}%
  \BibitemOpen
  \bibfield  {author} {\bibinfo {author} {\bibfnamefont {C.}~\bibnamefont
  {Lyu}}, \bibinfo {author} {\bibfnamefont {V.}~\bibnamefont {Montenegro}},\
  and\ \bibinfo {author} {\bibfnamefont {A.}~\bibnamefont {Bayat}},\ }\bibfield
   {title} {\bibinfo {title} {Accelerated variational algorithms for digital
  quantum simulation of many-body ground states},\ }\href
  {https://doi.org/10.22331/q-2020-09-16-324} {\bibfield  {journal} {\bibinfo
  {journal} {Quantum}\ }\textbf {\bibinfo {volume} {4}},\ \bibinfo {pages}
  {324} (\bibinfo {year} {2020})}\BibitemShut {NoStop}%
\bibitem [{\citenamefont {Chen}\ \emph {et~al.}(2010)\citenamefont {Chen},
  \citenamefont {Gu},\ and\ \citenamefont {Wen}}]{chen2010local}%
  \BibitemOpen
  \bibfield  {author} {\bibinfo {author} {\bibfnamefont {X.}~\bibnamefont
  {Chen}}, \bibinfo {author} {\bibfnamefont {Z.-C.}\ \bibnamefont {Gu}},\ and\
  \bibinfo {author} {\bibfnamefont {X.-G.}\ \bibnamefont {Wen}},\ }\bibfield
  {title} {\bibinfo {title} {Local unitary transformation, long-range quantum
  entanglement, wave function renormalization, and topological order},\ }\href
  {https://doi.org/10.1103/PhysRevB.82.155138} {\bibfield  {journal} {\bibinfo
  {journal} {Phys. Rev. B}\ }\textbf {\bibinfo {volume} {82}},\ \bibinfo
  {pages} {155138} (\bibinfo {year} {2010})}\BibitemShut {NoStop}%
\bibitem [{\citenamefont {Ciaramelletti}\ \emph {et~al.}(2025)\citenamefont
  {Ciaramelletti}, \citenamefont {Beseda}, \citenamefont {Consiglio},
  \citenamefont {Lepori}, \citenamefont {Apollaro},\ and\ \citenamefont
  {Paganelli}}]{ciaramelletti2025detecting}%
  \BibitemOpen
  \bibfield  {author} {\bibinfo {author} {\bibfnamefont {C.}~\bibnamefont
  {Ciaramelletti}}, \bibinfo {author} {\bibfnamefont {M.}~\bibnamefont
  {Beseda}}, \bibinfo {author} {\bibfnamefont {M.}~\bibnamefont {Consiglio}},
  \bibinfo {author} {\bibfnamefont {L.}~\bibnamefont {Lepori}}, \bibinfo
  {author} {\bibfnamefont {T.~J.~G.}\ \bibnamefont {Apollaro}},\ and\ \bibinfo
  {author} {\bibfnamefont {S.}~\bibnamefont {Paganelli}},\ }\bibfield  {title}
  {\bibinfo {title} {Detecting quasidegenerate ground states in topological
  models via the variational quantum eigensolver},\ }\href
  {https://doi.org/10.1103/PhysRevA.111.022437} {\bibfield  {journal} {\bibinfo
   {journal} {Phys. Rev. A}\ }\textbf {\bibinfo {volume} {111}},\ \bibinfo
  {pages} {022437} (\bibinfo {year} {2025})}\BibitemShut {NoStop}%
\bibitem [{\citenamefont {Sim}\ \emph {et~al.}(2019)\citenamefont {Sim},
  \citenamefont {Johnson},\ and\ \citenamefont
  {Aspuru-Guzik}}]{sim2019expressibility}%
  \BibitemOpen
  \bibfield  {author} {\bibinfo {author} {\bibfnamefont {S.}~\bibnamefont
  {Sim}}, \bibinfo {author} {\bibfnamefont {P.~D.}\ \bibnamefont {Johnson}},\
  and\ \bibinfo {author} {\bibfnamefont {A.}~\bibnamefont {Aspuru-Guzik}},\
  }\bibfield  {title} {\bibinfo {title} {Expressibility and entangling
  capability of parameterized quantum circuits for hybrid quantum-classical
  algorithms},\ }\href {https://doi.org/10.1002/qute.201900070} {\bibfield
  {journal} {\bibinfo  {journal} {Adv. Quantum Technol.}\ }\textbf {\bibinfo
  {volume} {2}},\ \bibinfo {pages} {1900070} (\bibinfo {year}
  {2019})}\BibitemShut {NoStop}%
\bibitem [{\citenamefont {Tang}\ \emph {et~al.}(2021)\citenamefont {Tang},
  \citenamefont {Shkolnikov}, \citenamefont {Barron}, \citenamefont {Grimsley},
  \citenamefont {Mayhall}, \citenamefont {Barnes},\ and\ \citenamefont
  {Economou}}]{tang2021qubit}%
  \BibitemOpen
  \bibfield  {author} {\bibinfo {author} {\bibfnamefont {H.~L.}\ \bibnamefont
  {Tang}}, \bibinfo {author} {\bibfnamefont {V.~O.}\ \bibnamefont
  {Shkolnikov}}, \bibinfo {author} {\bibfnamefont {G.~S.}\ \bibnamefont
  {Barron}}, \bibinfo {author} {\bibfnamefont {H.~R.}\ \bibnamefont
  {Grimsley}}, \bibinfo {author} {\bibfnamefont {N.~J.}\ \bibnamefont
  {Mayhall}}, \bibinfo {author} {\bibfnamefont {E.}~\bibnamefont {Barnes}},\
  and\ \bibinfo {author} {\bibfnamefont {S.~E.}\ \bibnamefont {Economou}},\
  }\bibfield  {title} {\bibinfo {title} {Qubit-{ADAPT-VQE}: {An} adaptive
  algorithm for constructing hardware-efficient ans{\"a}tze on a quantum
  processor},\ }\href {https://doi.org/10.1103/PRXQuantum.2.020310} {\bibfield
  {journal} {\bibinfo  {journal} {PRX Quantum}\ }\textbf {\bibinfo {volume}
  {2}},\ \bibinfo {pages} {020310} (\bibinfo {year} {2021})}\BibitemShut
  {NoStop}%
\bibitem [{\citenamefont {Du}\ \emph {et~al.}(2020)\citenamefont {Du},
  \citenamefont {Hsieh}, \citenamefont {Liu},\ and\ \citenamefont
  {Tao}}]{du2020expressive}%
  \BibitemOpen
  \bibfield  {author} {\bibinfo {author} {\bibfnamefont {Y.}~\bibnamefont
  {Du}}, \bibinfo {author} {\bibfnamefont {M.-H.}\ \bibnamefont {Hsieh}},
  \bibinfo {author} {\bibfnamefont {T.}~\bibnamefont {Liu}},\ and\ \bibinfo
  {author} {\bibfnamefont {D.}~\bibnamefont {Tao}},\ }\bibfield  {title}
  {\bibinfo {title} {Expressive power of parametrized quantum circuits},\
  }\href {https://doi.org/10.1103/PhysRevResearch.2.033125} {\bibfield
  {journal} {\bibinfo  {journal} {Phys. Rev. Res.}\ }\textbf {\bibinfo {volume}
  {2}},\ \bibinfo {pages} {033125} (\bibinfo {year} {2020})}\BibitemShut
  {NoStop}%
\bibitem [{\citenamefont {McClean}\ \emph {et~al.}(2018)\citenamefont
  {McClean}, \citenamefont {Boixo}, \citenamefont {Smelyanskiy}, \citenamefont
  {Babbush},\ and\ \citenamefont {Neven}}]{mcclean2018barren}%
  \BibitemOpen
  \bibfield  {author} {\bibinfo {author} {\bibfnamefont {J.~R.}\ \bibnamefont
  {McClean}}, \bibinfo {author} {\bibfnamefont {S.}~\bibnamefont {Boixo}},
  \bibinfo {author} {\bibfnamefont {V.~N.}\ \bibnamefont {Smelyanskiy}},
  \bibinfo {author} {\bibfnamefont {R.}~\bibnamefont {Babbush}},\ and\ \bibinfo
  {author} {\bibfnamefont {H.}~\bibnamefont {Neven}},\ }\bibfield  {title}
  {\bibinfo {title} {Barren plateaus in quantum neural network training
  landscapes},\ }\href {https://doi.org/10.1038/s41467-018-07090-4} {\bibfield
  {journal} {\bibinfo  {journal} {Nat. Commun.}\ }\textbf {\bibinfo {volume}
  {9}},\ \bibinfo {pages} {4812} (\bibinfo {year} {2018})}\BibitemShut
  {NoStop}%
\bibitem [{\citenamefont {Homeier}\ \emph
  {et~al.}(2021{\natexlab{a}})\citenamefont {Homeier}, \citenamefont
  {Schweizer}, \citenamefont {Aidelsburger}, \citenamefont {Fedorov},\ and\
  \citenamefont {Grusdt}}]{exp}%
  \BibitemOpen
  \bibfield  {author} {\bibinfo {author} {\bibfnamefont {L.}~\bibnamefont
  {Homeier}}, \bibinfo {author} {\bibfnamefont {C.}~\bibnamefont {Schweizer}},
  \bibinfo {author} {\bibfnamefont {M.}~\bibnamefont {Aidelsburger}}, \bibinfo
  {author} {\bibfnamefont {A.}~\bibnamefont {Fedorov}},\ and\ \bibinfo {author}
  {\bibfnamefont {F.}~\bibnamefont {Grusdt}},\ }\bibfield  {title} {\bibinfo
  {title} {${\mathbb{z}}_{2}$ lattice gauge theories and {Kitaev's} toric code:
  {A} scheme for analog quantum simulation},\ }\href
  {https://doi.org/10.1103/PhysRevB.104.085138} {\bibfield  {journal} {\bibinfo
   {journal} {Phys. Rev. B}\ }\textbf {\bibinfo {volume} {104}},\ \bibinfo
  {pages} {085138} (\bibinfo {year} {2021}{\natexlab{a}})}\BibitemShut
  {NoStop}%
\bibitem [{\citenamefont {Homeier}\ \emph
  {et~al.}(2021{\natexlab{b}})\citenamefont {Homeier}, \citenamefont
  {Schweizer}, \citenamefont {Aidelsburger}, \citenamefont {Fedorov},\ and\
  \citenamefont {Grusdt}}]{homeier2021z}%
  \BibitemOpen
  \bibfield  {author} {\bibinfo {author} {\bibfnamefont {L.}~\bibnamefont
  {Homeier}}, \bibinfo {author} {\bibfnamefont {C.}~\bibnamefont {Schweizer}},
  \bibinfo {author} {\bibfnamefont {M.}~\bibnamefont {Aidelsburger}}, \bibinfo
  {author} {\bibfnamefont {A.}~\bibnamefont {Fedorov}},\ and\ \bibinfo {author}
  {\bibfnamefont {F.}~\bibnamefont {Grusdt}},\ }\bibfield  {title} {\bibinfo
  {title} {Z 2 lattice gauge theories and kitaev's toric code: A scheme for
  analog quantum simulation},\ }\href
  {https://doi.org/10.1103/PhysRevB.104.085138} {\bibfield  {journal} {\bibinfo
   {journal} {Phys. Revi. B}\ }\textbf {\bibinfo {volume} {104}},\ \bibinfo
  {pages} {085138} (\bibinfo {year} {2021}{\natexlab{b}})}\BibitemShut
  {NoStop}%
\bibitem [{\citenamefont {Dennis}\ \emph {et~al.}(2002)\citenamefont {Dennis},
  \citenamefont {Kitaev}, \citenamefont {Landahl},\ and\ \citenamefont
  {Preskill}}]{dennis2002}%
  \BibitemOpen
  \bibfield  {author} {\bibinfo {author} {\bibfnamefont {E.}~\bibnamefont
  {Dennis}}, \bibinfo {author} {\bibfnamefont {A.}~\bibnamefont {Kitaev}},
  \bibinfo {author} {\bibfnamefont {A.}~\bibnamefont {Landahl}},\ and\ \bibinfo
  {author} {\bibfnamefont {J.}~\bibnamefont {Preskill}},\ }\bibfield  {title}
  {\bibinfo {title} {Topological quantum memory},\ }\href
  {https://doi.org/10.1063/1.1499754} {\bibfield  {journal} {\bibinfo
  {journal} {J. Math. Phys.}\ }\textbf {\bibinfo {volume} {43}},\ \bibinfo
  {pages} {4452} (\bibinfo {year} {2002})}\BibitemShut {NoStop}%
\bibitem [{\citenamefont {Levin}\ and\ \citenamefont {Wen}(2005)}]{Levin}%
  \BibitemOpen
  \bibfield  {author} {\bibinfo {author} {\bibfnamefont {M.~A.}\ \bibnamefont
  {Levin}}\ and\ \bibinfo {author} {\bibfnamefont {X.-G.}\ \bibnamefont
  {Wen}},\ }\bibfield  {title} {\bibinfo {title} {String-net condensation: {A}
  physical mechanism for topological phases},\ }\href
  {https://doi.org/10.1103/PhysRevB.71.045110} {\bibfield  {journal} {\bibinfo
  {journal} {Phys. Rev. B}\ }\textbf {\bibinfo {volume} {71}},\ \bibinfo
  {pages} {045110} (\bibinfo {year} {2005})}\BibitemShut {NoStop}%
\bibitem [{\citenamefont {Bravyi}\ \emph {et~al.}(2010)\citenamefont {Bravyi},
  \citenamefont {Hastings},\ and\ \citenamefont
  {Michalakis}}]{bravyi2010topological}%
  \BibitemOpen
  \bibfield  {author} {\bibinfo {author} {\bibfnamefont {S.}~\bibnamefont
  {Bravyi}}, \bibinfo {author} {\bibfnamefont {M.~B.}\ \bibnamefont
  {Hastings}},\ and\ \bibinfo {author} {\bibfnamefont {S.}~\bibnamefont
  {Michalakis}},\ }\bibfield  {title} {\bibinfo {title} {Topological quantum
  order: {Stability} under local perturbations},\ }\href
  {https://doi.org/10.1063/1.3490195} {\bibfield  {journal} {\bibinfo
  {journal} {J. Math. Phys.}\ }\textbf {\bibinfo {volume} {51}},\ \bibinfo
  {pages} {093512} (\bibinfo {year} {2010})}\BibitemShut {NoStop}%
\bibitem [{\citenamefont {Araujo~de Resende}(2020)}]{araujoderesende2020}%
  \BibitemOpen
  \bibfield  {author} {\bibinfo {author} {\bibfnamefont {M.~F.}\ \bibnamefont
  {Araujo~de Resende}},\ }\bibfield  {title} {\bibinfo {title} {A pedagogical
  overview on {2D} and {3D} {Toric Codes} and the origin of their topological
  orders},\ }\href {https://doi.org/10.1142/S0129055X20300022} {\bibfield
  {journal} {\bibinfo  {journal} {Rev. Math. Phys.}\ }\textbf {\bibinfo
  {volume} {32}},\ \bibinfo {pages} {2030002} (\bibinfo {year}
  {2020})}\BibitemShut {NoStop}%
\bibitem [{\citenamefont {Iqbal}\ \emph {et~al.}(2024)\citenamefont {Iqbal},
  \citenamefont {Tantivasukij}, \citenamefont {Verresen}, \citenamefont
  {Campbell}, \citenamefont {Dreiling}, \citenamefont {Figgatt}, \citenamefont
  {Gaebler}, \citenamefont {Johansen}, \citenamefont {Mills}, \citenamefont
  {Moses} \emph {et~al.}}]{iqbal2024topological}%
  \BibitemOpen
  \bibfield  {author} {\bibinfo {author} {\bibfnamefont {M.}~\bibnamefont
  {Iqbal}}, \bibinfo {author} {\bibfnamefont {N.}~\bibnamefont {Tantivasukij}},
  \bibinfo {author} {\bibfnamefont {R.}~\bibnamefont {Verresen}}, \bibinfo
  {author} {\bibfnamefont {S.~L.}\ \bibnamefont {Campbell}}, \bibinfo {author}
  {\bibfnamefont {J.~M.}\ \bibnamefont {Dreiling}}, \bibinfo {author}
  {\bibfnamefont {C.}~\bibnamefont {Figgatt}}, \bibinfo {author} {\bibfnamefont
  {J.~P.}\ \bibnamefont {Gaebler}}, \bibinfo {author} {\bibfnamefont
  {J.}~\bibnamefont {Johansen}}, \bibinfo {author} {\bibfnamefont
  {M.}~\bibnamefont {Mills}}, \bibinfo {author} {\bibfnamefont {S.~A.}\
  \bibnamefont {Moses}}, \emph {et~al.},\ }\bibfield  {title} {\bibinfo {title}
  {Topological order from measurements and feed-forward on a quantum
  processor},\ }\href {https://doi.org/10.1038/s42005-024-01698-3} {\bibfield
  {journal} {\bibinfo  {journal} {Commun. Phys.}\ }\textbf {\bibinfo {volume}
  {7}},\ \bibinfo {pages} {205} (\bibinfo {year} {2024})}\BibitemShut {NoStop}%
\bibitem [{\citenamefont {Aktar}\ \emph {et~al.}(2026)\citenamefont {Aktar},
  \citenamefont {Bhardwaj}, \citenamefont {B{\"a}rtschi}, \citenamefont
  {Bhattacharya},\ and\ \citenamefont {Eidenbenz}}]{aktar2026quantum}%
  \BibitemOpen
  \bibfield  {author} {\bibinfo {author} {\bibfnamefont {S.}~\bibnamefont
  {Aktar}}, \bibinfo {author} {\bibfnamefont {R.}~\bibnamefont {Bhardwaj}},
  \bibinfo {author} {\bibfnamefont {A.}~\bibnamefont {B{\"a}rtschi}}, \bibinfo
  {author} {\bibfnamefont {T.}~\bibnamefont {Bhattacharya}},\ and\ \bibinfo
  {author} {\bibfnamefont {S.}~\bibnamefont {Eidenbenz}},\ }\bibfield  {title}
  {\bibinfo {title} {Quantum data learning of topological-to-ferromagnetic
  phase transitions in the $2+1\text{D}$ toric code loop gas model},\ }\href
  {https://doi.org/10.1103/4d35-kw5d} {\bibfield  {journal} {\bibinfo
  {journal} {Phys. Rev. D}\ }\textbf {\bibinfo {volume} {113}},\ \bibinfo
  {pages} {114523} (\bibinfo {year} {2026})}\BibitemShut {NoStop}%
\bibitem [{\citenamefont {Chen}\ \emph {et~al.}(2024)\citenamefont {Chen},
  \citenamefont {Yan},\ and\ \citenamefont {Cui}}]{chen2024quantumcircuits}%
  \BibitemOpen
  \bibfield  {author} {\bibinfo {author} {\bibfnamefont {P.}~\bibnamefont
  {Chen}}, \bibinfo {author} {\bibfnamefont {B.}~\bibnamefont {Yan}},\ and\
  \bibinfo {author} {\bibfnamefont {S.~X.}\ \bibnamefont {Cui}},\ }\bibfield
  {title} {\bibinfo {title} {Quantum circuits for toric code and {X}-cube
  fracton model},\ }\href {https://doi.org/10.22331/q-2024-03-13-1276}
  {\bibfield  {journal} {\bibinfo  {journal} {Quantum}\ }\textbf {\bibinfo
  {volume} {8}},\ \bibinfo {pages} {1276} (\bibinfo {year} {2024})}\BibitemShut
  {NoStop}%
\bibitem [{\citenamefont {Liu}\ \emph {et~al.}(2022)\citenamefont {Liu},
  \citenamefont {Shtengel}, \citenamefont {Smith},\ and\ \citenamefont
  {Pollmann}}]{liu2022methods}%
  \BibitemOpen
  \bibfield  {author} {\bibinfo {author} {\bibfnamefont {Y.-J.}\ \bibnamefont
  {Liu}}, \bibinfo {author} {\bibfnamefont {K.}~\bibnamefont {Shtengel}},
  \bibinfo {author} {\bibfnamefont {A.}~\bibnamefont {Smith}},\ and\ \bibinfo
  {author} {\bibfnamefont {F.}~\bibnamefont {Pollmann}},\ }\bibfield  {title}
  {\bibinfo {title} {Methods for simulating string-net states and anyons on a
  digital quantum computer},\ }\href
  {https://doi.org/10.1103/PRXQuantum.3.040315} {\bibfield  {journal} {\bibinfo
   {journal} {PRX Quantum}\ }\textbf {\bibinfo {volume} {3}},\ \bibinfo {pages}
  {040315} (\bibinfo {year} {2022})}\BibitemShut {NoStop}%
\bibitem [{\citenamefont {Fowler}\ \emph {et~al.}(2012)\citenamefont {Fowler},
  \citenamefont {Mariantoni}, \citenamefont {Martinis},\ and\ \citenamefont
  {Cleland}}]{fowler2012surface}%
  \BibitemOpen
  \bibfield  {author} {\bibinfo {author} {\bibfnamefont {A.~G.}\ \bibnamefont
  {Fowler}}, \bibinfo {author} {\bibfnamefont {M.}~\bibnamefont {Mariantoni}},
  \bibinfo {author} {\bibfnamefont {J.~M.}\ \bibnamefont {Martinis}},\ and\
  \bibinfo {author} {\bibfnamefont {A.~N.}\ \bibnamefont {Cleland}},\
  }\bibfield  {title} {\bibinfo {title} {Surface codes: Towards practical
  large-scale quantum computation},\ }\href
  {https://doi.org/10.1103/PhysRevA.86.032324} {\bibfield  {journal} {\bibinfo
  {journal} {Phys. Rev. A}\ }\textbf {\bibinfo {volume} {86}},\ \bibinfo
  {pages} {032324} (\bibinfo {year} {2012})}\BibitemShut {NoStop}%
\bibitem [{\citenamefont {Piroli}\ \emph {et~al.}(2021)\citenamefont {Piroli},
  \citenamefont {Styliaris},\ and\ \citenamefont {Cirac}}]{piroli2021quantum}%
  \BibitemOpen
  \bibfield  {author} {\bibinfo {author} {\bibfnamefont {L.}~\bibnamefont
  {Piroli}}, \bibinfo {author} {\bibfnamefont {G.}~\bibnamefont {Styliaris}},\
  and\ \bibinfo {author} {\bibfnamefont {J.~I.}\ \bibnamefont {Cirac}},\
  }\bibfield  {title} {\bibinfo {title} {Quantum circuits assisted by local
  operations and classical communication: Transformations and phases of
  matter},\ }\href {https://doi.org/10.1103/PhysRevLett.127.220503} {\bibfield
  {journal} {\bibinfo  {journal} {Phys. Rev. Lett}\ }\textbf {\bibinfo {volume}
  {127}},\ \bibinfo {pages} {220503} (\bibinfo {year} {2021})}\BibitemShut
  {NoStop}%
\bibitem [{\citenamefont {Satzinger}\ \emph {et~al.}(2021)\citenamefont
  {Satzinger}, \citenamefont {Liu}, \citenamefont {Smith}, \citenamefont
  {Knapp}, \citenamefont {Newman}, \citenamefont {Jones}, \citenamefont {Chen},
  \citenamefont {Quintana}, \citenamefont {Mi}, \citenamefont {Dunsworth} \emph
  {et~al.}}]{satzinger2021realizing}%
  \BibitemOpen
  \bibfield  {author} {\bibinfo {author} {\bibfnamefont {K.~J.}\ \bibnamefont
  {Satzinger}}, \bibinfo {author} {\bibfnamefont {Y.}~\bibnamefont {Liu}},
  \bibinfo {author} {\bibfnamefont {A.}~\bibnamefont {Smith}}, \bibinfo
  {author} {\bibfnamefont {C.}~\bibnamefont {Knapp}}, \bibinfo {author}
  {\bibfnamefont {M.}~\bibnamefont {Newman}}, \bibinfo {author} {\bibfnamefont
  {C.}~\bibnamefont {Jones}}, \bibinfo {author} {\bibfnamefont
  {Z.}~\bibnamefont {Chen}}, \bibinfo {author} {\bibfnamefont {C.}~\bibnamefont
  {Quintana}}, \bibinfo {author} {\bibfnamefont {X.}~\bibnamefont {Mi}},
  \bibinfo {author} {\bibfnamefont {A.}~\bibnamefont {Dunsworth}}, \emph
  {et~al.},\ }\bibfield  {title} {\bibinfo {title} {Realizing topologically
  ordered states on a quantum processor},\ }\href
  {https://doi.org/10.1126/science.abi8378} {\bibfield  {journal} {\bibinfo
  {journal} {Science}\ }\textbf {\bibinfo {volume} {374}},\ \bibinfo {pages}
  {1237} (\bibinfo {year} {2021})}\BibitemShut {NoStop}%
\bibitem [{\citenamefont {Trebst}\ \emph {et~al.}(2007)\citenamefont {Trebst},
  \citenamefont {Werner}, \citenamefont {Troyer}, \citenamefont {Shtengel},\
  and\ \citenamefont {Nayak}}]{trebst2007breakdown}%
  \BibitemOpen
  \bibfield  {author} {\bibinfo {author} {\bibfnamefont {S.}~\bibnamefont
  {Trebst}}, \bibinfo {author} {\bibfnamefont {P.}~\bibnamefont {Werner}},
  \bibinfo {author} {\bibfnamefont {M.}~\bibnamefont {Troyer}}, \bibinfo
  {author} {\bibfnamefont {K.}~\bibnamefont {Shtengel}},\ and\ \bibinfo
  {author} {\bibfnamefont {C.}~\bibnamefont {Nayak}},\ }\bibfield  {title}
  {\bibinfo {title} {Breakdown of a topological phase: Quantum phase transition
  in a loop gas model with tension},\ }\href
  {https://doi.org/10.1103/PhysRevLett.98.070602} {\bibfield  {journal}
  {\bibinfo  {journal} {Phys. Rev. Lett.}\ }\textbf {\bibinfo {volume} {98}},\
  \bibinfo {pages} {070602} (\bibinfo {year} {2007})}\BibitemShut {NoStop}%
\bibitem [{\citenamefont {Dusuel}\ \emph {et~al.}(2011)\citenamefont {Dusuel},
  \citenamefont {Kamfor}, \citenamefont {Or{\'u}s}, \citenamefont {Schmidt},\
  and\ \citenamefont {Vidal}}]{dusuel2011robustness}%
  \BibitemOpen
  \bibfield  {author} {\bibinfo {author} {\bibfnamefont {S.}~\bibnamefont
  {Dusuel}}, \bibinfo {author} {\bibfnamefont {M.}~\bibnamefont {Kamfor}},
  \bibinfo {author} {\bibfnamefont {R.}~\bibnamefont {Or{\'u}s}}, \bibinfo
  {author} {\bibfnamefont {K.~P.}\ \bibnamefont {Schmidt}},\ and\ \bibinfo
  {author} {\bibfnamefont {J.}~\bibnamefont {Vidal}},\ }\bibfield  {title}
  {\bibinfo {title} {Robustness of a perturbed topological phase},\ }\href
  {https://doi.org/10.1103/PhysRevLett.106.107203} {\bibfield  {journal}
  {\bibinfo  {journal} {Phys. Rev. Lett.}\ }\textbf {\bibinfo {volume} {106}},\
  \bibinfo {pages} {107203} (\bibinfo {year} {2011})}\BibitemShut {NoStop}%
\bibitem [{\citenamefont {Hamma}\ \emph
  {et~al.}(2005{\natexlab{a}})\citenamefont {Hamma}, \citenamefont
  {Ionicioiu},\ and\ \citenamefont {Zanardi}}]{hamma2005topological}%
  \BibitemOpen
  \bibfield  {author} {\bibinfo {author} {\bibfnamefont {A.}~\bibnamefont
  {Hamma}}, \bibinfo {author} {\bibfnamefont {R.}~\bibnamefont {Ionicioiu}},\
  and\ \bibinfo {author} {\bibfnamefont {P.}~\bibnamefont {Zanardi}},\
  }\bibfield  {title} {\bibinfo {title} {Bipartite entanglement and entropic
  boundary law in lattice spin systems},\ }\href
  {https://doi.org/10.1103/PhysRevA.71.022315} {\bibfield  {journal} {\bibinfo
  {journal} {Phys. Rev. A}\ }\textbf {\bibinfo {volume} {71}},\ \bibinfo
  {pages} {022315} (\bibinfo {year} {2005}{\natexlab{a}})}\BibitemShut
  {NoStop}%
\bibitem [{\citenamefont {Zanardi}\ and\ \citenamefont
  {Paunkovi{\'c}}(2006)}]{zanardi2006ground}%
  \BibitemOpen
  \bibfield  {author} {\bibinfo {author} {\bibfnamefont {P.}~\bibnamefont
  {Zanardi}}\ and\ \bibinfo {author} {\bibfnamefont {N.}~\bibnamefont
  {Paunkovi{\'c}}},\ }\bibfield  {title} {\bibinfo {title} {Ground state
  overlap and quantum phase transitions},\ }\href
  {https://doi.org/10.1103/PhysRevE.74.031123} {\bibfield  {journal} {\bibinfo
  {journal} {Phys. Rev. E}\ }\textbf {\bibinfo {volume} {74}},\ \bibinfo
  {pages} {031123} (\bibinfo {year} {2006})}\BibitemShut {NoStop}%
\bibitem [{\citenamefont {Zarei}(2019)}]{zarei2019ising}%
  \BibitemOpen
  \bibfield  {author} {\bibinfo {author} {\bibfnamefont {M.~H.}\ \bibnamefont
  {Zarei}},\ }\bibfield  {title} {\bibinfo {title} {{Ising} order parameter and
  topological phase transitions: {Toric} code in a uniform magnetic field},\
  }\href {https://doi.org/10.1103/PhysRevB.100.125159} {\bibfield  {journal}
  {\bibinfo  {journal} {Phys. Rev. B}\ }\textbf {\bibinfo {volume} {100}},\
  \bibinfo {pages} {125159} (\bibinfo {year} {2019})}\BibitemShut {NoStop}%
\bibitem [{\citenamefont {Sun}\ \emph {et~al.}(2023)\citenamefont {Sun},
  \citenamefont {Shirakawa},\ and\ \citenamefont
  {Yunoki}}]{sun2023parametrized}%
  \BibitemOpen
  \bibfield  {author} {\bibinfo {author} {\bibfnamefont {R.-Y.}\ \bibnamefont
  {Sun}}, \bibinfo {author} {\bibfnamefont {T.}~\bibnamefont {Shirakawa}},\
  and\ \bibinfo {author} {\bibfnamefont {S.}~\bibnamefont {Yunoki}},\
  }\bibfield  {title} {\bibinfo {title} {Parametrized quantum circuit for
  weight-adjustable quantum loop gas},\ }\href
  {https://doi.org/10.1103/PhysRevB.107.L041109} {\bibfield  {journal}
  {\bibinfo  {journal} {Phys. Rev. B}\ }\textbf {\bibinfo {volume} {107}},\
  \bibinfo {pages} {L041109} (\bibinfo {year} {2023})}\BibitemShut {NoStop}%
\bibitem [{\citenamefont {Huang}\ \emph {et~al.}(2020)\citenamefont {Huang},
  \citenamefont {Kueng},\ and\ \citenamefont {Preskill}}]{huang2020predicting}%
  \BibitemOpen
  \bibfield  {author} {\bibinfo {author} {\bibfnamefont {H.-Y.}\ \bibnamefont
  {Huang}}, \bibinfo {author} {\bibfnamefont {R.}~\bibnamefont {Kueng}},\ and\
  \bibinfo {author} {\bibfnamefont {J.}~\bibnamefont {Preskill}},\ }\bibfield
  {title} {\bibinfo {title} {Predicting many properties of a quantum system
  from very few measurements},\ }\href
  {https://doi.org/10.1038/s41567-020-0932-7} {\bibfield  {journal} {\bibinfo
  {journal} {Nat. Phys.}\ }\textbf {\bibinfo {volume} {16}},\ \bibinfo {pages}
  {1050} (\bibinfo {year} {2020})}\BibitemShut {NoStop}%
\bibitem [{\citenamefont {Bravyi}\ \emph {et~al.}(2021)\citenamefont {Bravyi},
  \citenamefont {Sheldon}, \citenamefont {Kandala}, \citenamefont {Mckay},\
  and\ \citenamefont {Gambetta}}]{bravyi2021mitigating}%
  \BibitemOpen
  \bibfield  {author} {\bibinfo {author} {\bibfnamefont {S.}~\bibnamefont
  {Bravyi}}, \bibinfo {author} {\bibfnamefont {S.}~\bibnamefont {Sheldon}},
  \bibinfo {author} {\bibfnamefont {A.}~\bibnamefont {Kandala}}, \bibinfo
  {author} {\bibfnamefont {D.~C.}\ \bibnamefont {Mckay}},\ and\ \bibinfo
  {author} {\bibfnamefont {J.~M.}\ \bibnamefont {Gambetta}},\ }\bibfield
  {title} {\bibinfo {title} {Mitigating measurement errors in multiqubit
  systems},\ }\href {https://doi.org/10.1103/PhysRevA.103.042605} {\bibfield
  {journal} {\bibinfo  {journal} {Phys. Rev. A}\ }\textbf {\bibinfo {volume}
  {103}},\ \bibinfo {pages} {042605} (\bibinfo {year} {2021})}\BibitemShut
  {NoStop}%
\bibitem [{\citenamefont {Gard}\ \emph {et~al.}(2020)\citenamefont {Gard},
  \citenamefont {Zhu}, \citenamefont {Barron}, \citenamefont {Mayhall},
  \citenamefont {Economou},\ and\ \citenamefont {Barnes}}]{gard2020efficient}%
  \BibitemOpen
  \bibfield  {author} {\bibinfo {author} {\bibfnamefont {B.~T.}\ \bibnamefont
  {Gard}}, \bibinfo {author} {\bibfnamefont {L.}~\bibnamefont {Zhu}}, \bibinfo
  {author} {\bibfnamefont {G.~S.}\ \bibnamefont {Barron}}, \bibinfo {author}
  {\bibfnamefont {N.~J.}\ \bibnamefont {Mayhall}}, \bibinfo {author}
  {\bibfnamefont {S.~E.}\ \bibnamefont {Economou}},\ and\ \bibinfo {author}
  {\bibfnamefont {E.}~\bibnamefont {Barnes}},\ }\bibfield  {title} {\bibinfo
  {title} {Efficient symmetry-preserving state preparation circuits for the
  variational quantum eigensolver algorithm},\ }\href
  {https://doi.org/10.1038/s41534-019-0240-1} {\bibfield  {journal} {\bibinfo
  {journal} {npj Quantum Inf.}\ }\textbf {\bibinfo {volume} {6}},\ \bibinfo
  {pages} {10} (\bibinfo {year} {2020})}\BibitemShut {NoStop}%
\bibitem [{\citenamefont {Li}\ \emph {et~al.}(2024)\citenamefont {Li},
  \citenamefont {Huang}, \citenamefont {Hou}, \citenamefont {Li}, \citenamefont
  {Wang},\ and\ \citenamefont {Bayat}}]{li2024ensemble}%
  \BibitemOpen
  \bibfield  {author} {\bibinfo {author} {\bibfnamefont {Q.}~\bibnamefont
  {Li}}, \bibinfo {author} {\bibfnamefont {Y.}~\bibnamefont {Huang}}, \bibinfo
  {author} {\bibfnamefont {X.}~\bibnamefont {Hou}}, \bibinfo {author}
  {\bibfnamefont {Y.}~\bibnamefont {Li}}, \bibinfo {author} {\bibfnamefont
  {X.}~\bibnamefont {Wang}},\ and\ \bibinfo {author} {\bibfnamefont
  {A.}~\bibnamefont {Bayat}},\ }\bibfield  {title} {\bibinfo {title}
  {Ensemble-learning error mitigation for variational quantum shallow-circuit
  classifiers},\ }\href {https://doi.org/10.1103/PhysRevResearch.6.013027}
  {\bibfield  {journal} {\bibinfo  {journal} {Phys. Rev. Res}\ }\textbf
  {\bibinfo {volume} {6}},\ \bibinfo {pages} {013027} (\bibinfo {year}
  {2024})}\BibitemShut {NoStop}%
\bibitem [{\citenamefont {Sweke}\ \emph {et~al.}(2020)\citenamefont {Sweke},
  \citenamefont {Wilde}, \citenamefont {Meyer}, \citenamefont {Schuld},
  \citenamefont {Faehrmann}, \citenamefont {Meynard-Piganeau},\ and\
  \citenamefont {Eisert}}]{sweke2020stochastic}%
  \BibitemOpen
  \bibfield  {author} {\bibinfo {author} {\bibfnamefont {R.}~\bibnamefont
  {Sweke}}, \bibinfo {author} {\bibfnamefont {F.}~\bibnamefont {Wilde}},
  \bibinfo {author} {\bibfnamefont {J.}~\bibnamefont {Meyer}}, \bibinfo
  {author} {\bibfnamefont {M.}~\bibnamefont {Schuld}}, \bibinfo {author}
  {\bibfnamefont {P.~K.}\ \bibnamefont {Faehrmann}}, \bibinfo {author}
  {\bibfnamefont {B.}~\bibnamefont {Meynard-Piganeau}},\ and\ \bibinfo {author}
  {\bibfnamefont {J.}~\bibnamefont {Eisert}},\ }\bibfield  {title} {\bibinfo
  {title} {Stochastic gradient descent for hybrid quantum-classical
  optimization},\ }\href {https://doi.org/10.22331/q-2020-08-31-314} {\bibfield
   {journal} {\bibinfo  {journal} {Quantum}\ }\textbf {\bibinfo {volume} {4}},\
  \bibinfo {pages} {314} (\bibinfo {year} {2020})}\BibitemShut {NoStop}%
\bibitem [{\citenamefont {Abraham}\ \emph {et~al.}(2019)\citenamefont
  {Abraham}, \citenamefont {Akhalwaya} \emph {et~al.}}]{abraham2019qiskit}%
  \BibitemOpen
  \bibfield  {author} {\bibinfo {author} {\bibfnamefont {H.}~\bibnamefont
  {Abraham}}, \bibinfo {author} {\bibfnamefont {I.~Y.}\ \bibnamefont
  {Akhalwaya}}, \emph {et~al.},\ }\href
  {https://doi.org/10.5281/zenodo.2562110} {\bibinfo {title} {{Qiskit}: {An}
  open-source framework for quantum computing}},\ \bibinfo {howpublished}
  {Zenodo} (\bibinfo {year} {2019})\BibitemShut {NoStop}%
\bibitem [{\citenamefont {Hamma}\ \emph {et~al.}(2008)\citenamefont {Hamma},
  \citenamefont {Zhang}, \citenamefont {Haas},\ and\ \citenamefont
  {Lidar}}]{hamma2008entanglement}%
  \BibitemOpen
  \bibfield  {author} {\bibinfo {author} {\bibfnamefont {A.}~\bibnamefont
  {Hamma}}, \bibinfo {author} {\bibfnamefont {W.}~\bibnamefont {Zhang}},
  \bibinfo {author} {\bibfnamefont {S.}~\bibnamefont {Haas}},\ and\ \bibinfo
  {author} {\bibfnamefont {D.}~\bibnamefont {Lidar}},\ }\bibfield  {title}
  {\bibinfo {title} {Entanglement, fidelity, and topological entropy in a
  quantum phase transition to topological order},\ }\href
  {https://doi.org/10.1103/PhysRevB.77.155111} {\bibfield  {journal} {\bibinfo
  {journal} {Phys. Rev. B}\ }\textbf {\bibinfo {volume} {77}},\ \bibinfo
  {pages} {155111} (\bibinfo {year} {2008})}\BibitemShut {NoStop}%
\bibitem [{\citenamefont {Hamma}\ \emph
  {et~al.}(2005{\natexlab{b}})\citenamefont {Hamma}, \citenamefont
  {Ionicioiu},\ and\ \citenamefont {Zanardi}}]{hamma2005ground}%
  \BibitemOpen
  \bibfield  {author} {\bibinfo {author} {\bibfnamefont {A.}~\bibnamefont
  {Hamma}}, \bibinfo {author} {\bibfnamefont {R.}~\bibnamefont {Ionicioiu}},\
  and\ \bibinfo {author} {\bibfnamefont {P.}~\bibnamefont {Zanardi}},\
  }\bibfield  {title} {\bibinfo {title} {Ground state entanglement and
  geometric entropy in the {Kitaev} model},\ }\href
  {https://doi.org/10.1016/j.physleta.2005.01.060} {\bibfield  {journal}
  {\bibinfo  {journal} {Phys. Lett. A}\ }\textbf {\bibinfo {volume} {337}},\
  \bibinfo {pages} {22} (\bibinfo {year} {2005}{\natexlab{b}})}\BibitemShut
  {NoStop}%
\bibitem [{\citenamefont {Kitaev}\ and\ \citenamefont
  {Preskill}(2006)}]{kitaev2006topological}%
  \BibitemOpen
  \bibfield  {author} {\bibinfo {author} {\bibfnamefont {A.}~\bibnamefont
  {Kitaev}}\ and\ \bibinfo {author} {\bibfnamefont {J.}~\bibnamefont
  {Preskill}},\ }\bibfield  {title} {\bibinfo {title} {Topological entanglement
  entropy},\ }\href {https://doi.org/10.1103/PhysRevLett.96.110404} {\bibfield
  {journal} {\bibinfo  {journal} {Phys. Rev. Lett.}\ }\textbf {\bibinfo
  {volume} {96}},\ \bibinfo {pages} {110404} (\bibinfo {year}
  {2006})}\BibitemShut {NoStop}%
\bibitem [{\citenamefont {de~Oliveira}\ \emph
  {et~al.}(2006{\natexlab{a}})\citenamefont {de~Oliveira}, \citenamefont
  {Rigolin}, \citenamefont {de~Oliveira},\ and\ \citenamefont
  {Miranda}}]{de2006multipartite}%
  \BibitemOpen
  \bibfield  {author} {\bibinfo {author} {\bibfnamefont {T.~R.}\ \bibnamefont
  {de~Oliveira}}, \bibinfo {author} {\bibfnamefont {G.}~\bibnamefont
  {Rigolin}}, \bibinfo {author} {\bibfnamefont {M.~C.}\ \bibnamefont
  {de~Oliveira}},\ and\ \bibinfo {author} {\bibfnamefont {E.}~\bibnamefont
  {Miranda}},\ }\bibfield  {title} {\bibinfo {title} {Multipartite entanglement
  signature of quantum phase transitions},\ }\href
  {https://doi.org/10.1103/PhysRevLett.97.170401} {\bibfield  {journal}
  {\bibinfo  {journal} {Phys. Rev. lett}\ }\textbf {\bibinfo {volume} {97}},\
  \bibinfo {pages} {170401} (\bibinfo {year} {2006}{\natexlab{a}})}\BibitemShut
  {NoStop}%
\bibitem [{\citenamefont {de~Oliveira}\ \emph
  {et~al.}(2006{\natexlab{b}})\citenamefont {de~Oliveira}, \citenamefont
  {Rigolin}, \citenamefont {de~Oliveira},\ and\ \citenamefont
  {Rodriguez-Chavez}}]{de2006global}%
  \BibitemOpen
  \bibfield  {author} {\bibinfo {author} {\bibfnamefont {T.~R.}\ \bibnamefont
  {de~Oliveira}}, \bibinfo {author} {\bibfnamefont {G.}~\bibnamefont
  {Rigolin}}, \bibinfo {author} {\bibfnamefont {E.~L.}\ \bibnamefont
  {de~Oliveira}},\ and\ \bibinfo {author} {\bibfnamefont {E.~I.}\ \bibnamefont
  {Rodriguez-Chavez}},\ }\bibfield  {title} {\bibinfo {title} {Genuine
  multipartite entanglement in quantum phase transitions},\ }\href
  {https://doi.org/10.1103/PhysRevA.73.010305} {\bibfield  {journal} {\bibinfo
  {journal} {Phys. Rev. A}\ }\textbf {\bibinfo {volume} {73}},\ \bibinfo
  {pages} {010305(R)} (\bibinfo {year} {2006}{\natexlab{b}})}\BibitemShut
  {NoStop}%
\bibitem [{\citenamefont {Montakhab}\ and\ \citenamefont
  {Asadian}(2010)}]{montakhab2010multipartite}%
  \BibitemOpen
  \bibfield  {author} {\bibinfo {author} {\bibfnamefont {A.}~\bibnamefont
  {Montakhab}}\ and\ \bibinfo {author} {\bibfnamefont {A.}~\bibnamefont
  {Asadian}},\ }\bibfield  {title} {\bibinfo {title} {Multipartite entanglement
  and quantum phase transitions in the one-, two-, and three-dimensional
  transverse-field ising model},\ }\href
  {https://doi.org/10.1103/PhysRevA.82.062313} {\bibfield  {journal} {\bibinfo
  {journal} {Phys. Rev. A}\ }\textbf {\bibinfo {volume} {82}},\ \bibinfo
  {pages} {062313} (\bibinfo {year} {2010})}\BibitemShut {NoStop}%
\bibitem [{\citenamefont {Samimi}\ \emph {et~al.}(2022)\citenamefont {Samimi},
  \citenamefont {Zarei},\ and\ \citenamefont {Montakhab}}]{samimi2022global}%
  \BibitemOpen
  \bibfield  {author} {\bibinfo {author} {\bibfnamefont {E.}~\bibnamefont
  {Samimi}}, \bibinfo {author} {\bibfnamefont {M.~H.}\ \bibnamefont {Zarei}},\
  and\ \bibinfo {author} {\bibfnamefont {A.}~\bibnamefont {Montakhab}},\
  }\bibfield  {title} {\bibinfo {title} {Global entanglement in a topological
  quantum phase transition},\ }\href
  {https://doi.org/10.1103/PhysRevA.105.032438} {\bibfield  {journal} {\bibinfo
   {journal} {Phys. Rev. A}\ }\textbf {\bibinfo {volume} {105}},\ \bibinfo
  {pages} {032438} (\bibinfo {year} {2022})}\BibitemShut {NoStop}%
\bibitem [{\citenamefont {Samimi}\ \emph {et~al.}(2023)\citenamefont {Samimi},
  \citenamefont {Zarei},\ and\ \citenamefont
  {Montakhab}}]{samimi2023conditional}%
  \BibitemOpen
  \bibfield  {author} {\bibinfo {author} {\bibfnamefont {E.}~\bibnamefont
  {Samimi}}, \bibinfo {author} {\bibfnamefont {M.~H.}\ \bibnamefont {Zarei}},\
  and\ \bibinfo {author} {\bibfnamefont {A.}~\bibnamefont {Montakhab}},\
  }\bibfield  {title} {\bibinfo {title} {Conditional global entanglement in a
  kosterlitz-thouless quantum phase transition},\ }\href
  {https://doi.org/10.1103/PhysRevA.107.052412} {\bibfield  {journal} {\bibinfo
   {journal} {Phys. Rev. A}\ }\textbf {\bibinfo {volume} {107}},\ \bibinfo
  {pages} {052412} (\bibinfo {year} {2023})}\BibitemShut {NoStop}%
\bibitem [{\citenamefont {Barron}\ \emph {et~al.}(2021)\citenamefont {Barron},
  \citenamefont {Gard}, \citenamefont {Altman}, \citenamefont {Mayhall},
  \citenamefont {Barnes},\ and\ \citenamefont
  {Economou}}]{barron2021preserving}%
  \BibitemOpen
  \bibfield  {author} {\bibinfo {author} {\bibfnamefont {G.~S.}\ \bibnamefont
  {Barron}}, \bibinfo {author} {\bibfnamefont {B.~T.}\ \bibnamefont {Gard}},
  \bibinfo {author} {\bibfnamefont {O.~J.}\ \bibnamefont {Altman}}, \bibinfo
  {author} {\bibfnamefont {N.~J.}\ \bibnamefont {Mayhall}}, \bibinfo {author}
  {\bibfnamefont {E.}~\bibnamefont {Barnes}},\ and\ \bibinfo {author}
  {\bibfnamefont {S.~E.}\ \bibnamefont {Economou}},\ }\bibfield  {title}
  {\bibinfo {title} {Preserving symmetries for variational quantum eigensolvers
  in the presence of noise},\ }\href
  {https://doi.org/10.1103/PhysRevApplied.16.034003} {\bibfield  {journal}
  {\bibinfo  {journal} {Phys. Rev. Applied}\ }\textbf {\bibinfo {volume}
  {16}},\ \bibinfo {pages} {034003} (\bibinfo {year} {2021})}\BibitemShut
  {NoStop}%
\bibitem [{\citenamefont {Preskill}(2018)}]{preskill2018quantum}%
  \BibitemOpen
  \bibfield  {author} {\bibinfo {author} {\bibfnamefont {J.}~\bibnamefont
  {Preskill}},\ }\bibfield  {title} {\bibinfo {title} {Quantum computing in the
  {NISQ} era and beyond},\ }\href {https://doi.org/10.22331/q-2018-08-06-79}
  {\bibfield  {journal} {\bibinfo  {journal} {Quantum}\ }\textbf {\bibinfo
  {volume} {2}},\ \bibinfo {pages} {79} (\bibinfo {year} {2018})}\BibitemShut
  {NoStop}%
\bibitem [{\citenamefont {Wallman}\ \emph {et~al.}(2015)\citenamefont
  {Wallman}, \citenamefont {Granade}, \citenamefont {Harper},\ and\
  \citenamefont {Flammia}}]{wallman2016noise}%
  \BibitemOpen
  \bibfield  {author} {\bibinfo {author} {\bibfnamefont {J.~J.}\ \bibnamefont
  {Wallman}}, \bibinfo {author} {\bibfnamefont {C.}~\bibnamefont {Granade}},
  \bibinfo {author} {\bibfnamefont {R.}~\bibnamefont {Harper}},\ and\ \bibinfo
  {author} {\bibfnamefont {S.~T.}\ \bibnamefont {Flammia}},\ }\bibfield
  {title} {\bibinfo {title} {Estimating the coherence of noise},\ }\href
  {https://doi.org/10.1088/1367-2630/17/11/113020} {\bibfield  {journal}
  {\bibinfo  {journal} {New J. Phys.}\ }\textbf {\bibinfo {volume} {17}},\
  \bibinfo {pages} {113020} (\bibinfo {year} {2015})}\BibitemShut {NoStop}%
\bibitem [{\citenamefont {Aguado}\ and\ \citenamefont
  {Vidal}(2008)}]{vidal2008entanglement}%
  \BibitemOpen
  \bibfield  {author} {\bibinfo {author} {\bibfnamefont {M.}~\bibnamefont
  {Aguado}}\ and\ \bibinfo {author} {\bibfnamefont {G.}~\bibnamefont {Vidal}},\
  }\bibfield  {title} {\bibinfo {title} {Entanglement renormalization and
  topological order},\ }\href {https://doi.org/10.1103/PhysRevLett.100.070404}
  {\bibfield  {journal} {\bibinfo  {journal} {Phys. Rev. Lett.}\ }\textbf
  {\bibinfo {volume} {100}},\ \bibinfo {pages} {070404} (\bibinfo {year}
  {2008})}\BibitemShut {NoStop}%
\bibitem [{\citenamefont {Sachdev}(2011)}]{SachdevQuantumPhaseTransitions}%
  \BibitemOpen
  \bibfield  {author} {\bibinfo {author} {\bibfnamefont {S.}~\bibnamefont
  {Sachdev}},\ }\href {https://doi.org/10.1017/CBO9780511973765} {\emph
  {\bibinfo {title} {Quantum Phase Transitions}}},\ \bibinfo {edition} {2nd}\
  ed.\ (\bibinfo  {publisher} {Cambridge University Press},\ \bibinfo {address}
  {Cambridge},\ \bibinfo {year} {2011})\BibitemShut {NoStop}%
\bibitem [{\citenamefont {Meyer}\ and\ \citenamefont
  {Wallach}(2002)}]{MeyerWallach2002}%
  \BibitemOpen
  \bibfield  {author} {\bibinfo {author} {\bibfnamefont {D.~A.}\ \bibnamefont
  {Meyer}}\ and\ \bibinfo {author} {\bibfnamefont {N.~R.}\ \bibnamefont
  {Wallach}},\ }\bibfield  {title} {\bibinfo {title} {Global entanglement in
  multiparticle systems},\ }\href {https://doi.org/10.1063/1.1497700}
  {\bibfield  {journal} {\bibinfo  {journal} {J. Math. Phys.}\ }\textbf
  {\bibinfo {volume} {43}},\ \bibinfo {pages} {4273} (\bibinfo {year}
  {2002})}\BibitemShut {NoStop}%
\bibitem [{\citenamefont {Nielsen}\ and\ \citenamefont
  {Chuang}(2010)}]{NielsenChuang}%
  \BibitemOpen
  \bibfield  {author} {\bibinfo {author} {\bibfnamefont {M.~A.}\ \bibnamefont
  {Nielsen}}\ and\ \bibinfo {author} {\bibfnamefont {I.~L.}\ \bibnamefont
  {Chuang}},\ }\href@noop {} {\emph {\bibinfo {title} {Quantum Computation and
  Quantum Information}}},\ \bibinfo {edition} {10th}\ ed.\ (\bibinfo
  {publisher} {Cambridge University Press},\ \bibinfo {address} {Cambridge},\
  \bibinfo {year} {2010})\BibitemShut {NoStop}%
\end{thebibliography}%

\end{document}